\def\HST{{\it HST}}

\def\tcar{\futurelet\next\testnextcar}
\def\testnextcar{\ifhmode\ifcat\next.\else\ \fi\fi}
\def\kms{km~s$^{-1}$\tcar}

\def\etal{{et~al.}}

\def\QSO1038{{Q~1038$-$2712}}
\def\qso2343{Q~2343$+$125}
\def\pg2302{PG~2302$+$029}

\def\Atoday{\ifcase\month\or
  January\or February\or March\or April\or May\or June\or
  July\or August\or September\or October\or November\or December\fi
  \space\number\day, \number\year}

\def\Lya{\hbox{Ly-$\alpha$}}
\def\Lyb{\hbox{Ly-$\beta$}}
\def\Lyg{\hbox{Ly-$\gamma$}}
\def\Lyd{\hbox{Ly-$\delta$}}
\def\Lye{\hbox{Ly-$\epsilon$}}
\def\Lyz{\hbox{Ly-$\zeta$}}
\def\Lyeta{\hbox{Ly-$\eta$}}
\def\Lyth{\hbox{Ly-$\theta$}}
\def\Lyi{\hbox{Ly-$\iota$}}
\def\Lyk{\hbox{Ly-$\kappa$}}
\def\Lyl{\hbox{Ly-$\lambda$}}

\def\ll{\hbox{$\lambda\lambda$}}
\def\l{$\lambda$}

\def\papertwo{Paper~II}

\def\cycle1{Cycle~1}
\def\eg{{\it{e.g.,}}}
\def\zsearch{{\it ZSEARCH}}
\documentstyle[emulateapj]{article}
\begin{document}
\overfullrule=0pt
\input psfig

\title{THE HUBBLE SPACE TELESCOPE QUASAR ABSORPTION LINE KEY
PROJECT. XIII.\\
 A Census of Absorption Line Systems at Low Redshift\altaffilmark{1}}
\altaffiltext{1}{Based on observations with the
NASA/ESA {\it Hubble Space Telescope}, obtained at the Space Telescope 
Science Institute, which is operated by the Association of
Universities for Research in Astronomy, Inc., under NASA contract
NAS5-26555.}

\author{
Buell~T.~Jannuzi,\altaffilmark{2} \altaffiltext{2}{National Optical
Astronomy Observatories, P.O. Box 26732, Tucson, AZ~85719,
email:~jannuzi@noao.edu}
John~N.~Bahcall,\altaffilmark{3}\altaffiltext{3}{Institute for Advanced 
Study, School of Natural Sciences, Olden Lane, Princeton, NJ~08540,
email:~jnb@IAS.edu, sofia@IAS.edu} 
Jacqueline~Bergeron,\altaffilmark{4}\altaffiltext{4}{Institut~d'Astrophysique,
98 bis Boulevard Arago, F-75014 Paris, France, email:~bergeron@iap.fr}
Alec~Boksenberg,\altaffilmark{5}\altaffiltext{5}{Institute of Astronomy,
University of Cambridge, Madingley Road, Cambridge CB3 OHA, UK,
email:~boksy@ast.cam.ac.uk} 
George~F.~Hartig,\altaffilmark{6}\altaffiltext{6}{Space Telescope Science 
Institute, 3700~San Martin Drive, Baltimore, MD~21218,
email:~hartig@stsci.edu}  
Sofia~Kirhakos,\altaffilmark{3}
W.~L.~W.~Sargent,\altaffilmark{7}\altaffiltext{7}{Robinson Laboratory 
105-24, California Institute of Technology, Pasadena, CA 91125,
email:~wws@deimos.caltech.edu} 
Blair~D.~Savage,\altaffilmark{8}\altaffiltext{8}{Department of Astronomy, 
University of Wisconsin, 475 N. Charter Street, Madison, WI 53706,
email:~savage@uwast.astro.wisc.edu}  
Donald~P.~Schneider,\altaffilmark{9}\altaffiltext{9}{Department of
Astronomy and Astrophysics, The Pennsylvania State University,
University Park, PA 16802, email:~dps@astro.psu.edu}
David~A.~Turnshek,\altaffilmark{10}\altaffiltext{10}{Department of
Physics \& Astronomy, University of Pittsburgh, Pittsburgh, PA 15260,
email:~turnshek@vms.cis.pitt.edu} 
Ray~J.~Weymann,\altaffilmark{11}\altaffiltext{11}{The Observatories of
the Carnegie Institution of Washington, 813 Santa Barbara Street,
Pasadena, CA 91101, email:~rjw@ociw.edu} 
\& Arthur~M.~Wolfe\altaffilmark{12}\altaffiltext{12}{Center for
Astrophysics \& Space Sciences, C011, University of California San
Diego, La Jolla, CA 92093, email:~art@ucsd.edu\hfil\break
\hbox to\hsize{\hfill}} 
}

\begin{abstract}

We present a catalogue of absorption lines obtained from the analysis
of the ultra-violet spectra of 66 quasars.  The data were acquired
with the {\it Faint Object Spectrograph} of the {\it Hubble Space
Telescope} ({\it HST}) as part of the Quasar Absorption Line Survey, a
Key Project for the first four cycles of {\it HST} observations.  This
is the third of a series of catalogues of absorption lines produced
from the survey and increases the number of quasars whose higher
resolution ($R=1300$) spectra we have published from 17 to 83.  The
general properties and execution of the survey are reviewed, including
descriptions of the final sample of observed objects and the
algorithmic processes used to construct the catalogue.  This database
is suitable for a wide variety of studies of gaseous systems in the
nearby Universe.

This third catalogue includes 2594 absorption lines and brings the
total number of absorption lines in the combined catalogue to 3238.
The third catalogue has 878 identified \Lya~lines, 27 extensive metal
line systems (detected absorption lines from four or more metal ions),
88 C~IV systems, and 34 O~VI systems.  The combined catalogue contains
the following numbers of extragalactic absorption lines: 1,129
\Lya~lines, 107~C~IV systems, 41~O~VI systems, 16~Lyman$-$limit
systems, and one damped \Lya~system (in the spectrum of
PG~0935$+$416).  In addition, there are 25 pairs of identified
\Lya~lines that are candidate C~IV doublets. Of the 122 identified 
C~IV and candidate C~IV systems in the completely identified sample of
absorption lines, 24$\pm$5~are expected to be chance coincidences of
other lines (based upon Monte Carlo simulations).

The detection of a single damped \Lya~system in a path length of
$\Delta z = 49$~yields an observed number of damped systems per unit
redshift of $(dN/dz)_{\rm damp}(z=0.58)=0.020$ with 95\% confidence
boundaries of 0.001 to 0.096 systems per unit redshift.

We include notes on our analysis of each of the observed quasars and
the absorption systems detected in each spectrum.  Some especially
interesting systems include low redshift
\Lya~absorbers suitable for extensive follow-up observations (\eg~in
the spectra of TON~28 and PG~1216$+$069), possibly physically
associated pairs of extensive metal line absorption systems (\eg~in
the spectrum of PG~0117$+$213), and systems known to be associated
with galaxies (\eg~in the spectrum of 3C~232).

The spectra of five broad absorption line (BAL) quasars (UM~425,
PG~1254$+$047, PG~1411$+$442, PG~1700$+$518, and PG~2112$+$059) can be
found in this third catalogue, bringing the total number of BAL
quasars in the combined catalogue to six (including PG~0043$+$039).

\end{abstract}
\bigskip

\vfill\eject
\section{Introduction}
\label{sec-intro}

The {\it Hubble Space Telescope}~({\it HST}) and its spectrographs
have revolutionized the study of low redshift gaseous systems and
Galactic halo gas by providing the spectral resolution and sensitivity
needed to observe ultra-violet (UV) absorption lines in the spectra of
quasars.  Since the majority of the strong resonance absorption lines
of cosmically abundant ions, including many of the most powerful
diagnostics for the physical conditions of an absorber, occur in the
rest frame UV, the effective study of the nearby gaseous Universe
requires the use of UV spectroscopy (Spitzer 1956).

The primary goal of the Quasar Absorption Line Survey, a Key Project
during the first four cycles of observations with {\it HST}, was to
produce a large and homogeneous catalogue of absorbers suitable for
the study of gaseous systems at low redshifts.  The results from our
initial observations and analysis of the higher resolution ($R=1300$)
UV spectra of 17 quasars have been previously presented in our first
two catalogue papers (Bahcall \etal~1993a, Paper~I, but hereafter
CAT1; Bahcall \etal~1996, Paper~VII, but hereafter CAT2). The basic
survey design, data calibration, and analysis software are discussed
by Schneider \etal~(1993; hereafter \papertwo). In this paper (CAT3)
we present the rest of the UV spectra obtained as part of the Key
Project and also the Bahcall Guaranteed Time Observer (GTO) programs.
The total number of observed quasars in our sample is 92, 83
of which have been observed with one or more of the higher resolution
gratings.  Our analysis of these higher resolution spectra yield this
third and by far the largest of our catalogues of absorption lines. 
A series of companion papers present the results of the analyses of the
combined catalogue (CAT1, CAT2, and CAT3 taken together).

The paper is organized as follows: a description of the observational
properties of the survey including the final list of observed targets
and a review of the calibration of the spectra~(\S\ref{sec-obs}); a
review of how this third catalogue of absorption line systems was
constructed (\S\ref{sec-constructing}), including a description of
the algorithms used for the selection and measurement of the
absorption lines (\S\ref{sec-select}) and Lyman$-$limit
systems~(\S3.2), a summary of the current algorithms
for identifying absorption lines~(\S\ref{sec-identify}), which have
undergone some changes since CAT2, and our tests for the reliability of
our C~IV and O~VI identifications (\S\ref{sec-reliability}); line
identifications for the spectra presented in this paper including
comments on the identifications on an object-by-object
basis~(\S\ref{sec-notes}); a discussion of what the catalogue reveals
about various types of absorbers in~(\S\ref{sec-census}); and a brief
summary of this paper~(\S\ref{sec-summary}).

\section{Observations}
\label{sec-obs}

In this section we review the sample of quasars observed during the
survey, the instrumentation and techniques used for the observations,
and the calibration of the UV spectra.  The original design and
observational procedures used in the survey are described in detail in
\papertwo~and the first two catalogue papers (CAT1, CAT2).  Here we
review only the most important characteristics of the observed sample
and observing procedures.
\break
\subsection{Final Sample of Observed Objects}
\label{sec-sample}

The Quasar Absorption Line Survey was originally intended to be
comprised of observations of two samples of objects. Observations of
the ``primary sample'' were intended to provide the bulk of the
absorption lines in the final catalogue. These objects were selected
to have the following properties: Galactic latitude, $|b|$
$>20^{\circ}$; redshifts determined from slit spectroscopy which lie
between 0.15 and 2.0; well determined apparent optical magnitudes with
preference given to brighter targets. No reference was made to
existing UV data on a potential target during the initial sample
selection.  The radio properties of the quasars were also ignored
during selection of these targets.  The spectra obtained of the
primary sample were to be of sufficient spectral resolution ($R=1300$)
and signal-to-noise ratio (SNR$=$30) for our rest equivalent width
limits on absorption lines to be similar to those of existing large
ground-based surveys, and for important absorption line doublets
(\eg~C~IV separation of 2.57~\AA) to be well resolved.  To investigate
the strong, but rare, absorption systems such as damped
Lyman-$\alpha$~and Lyman$-$limit systems and to facilitate radio
spectroscopic follow-up of these systems, we constructed a second list
of targets with moderate redshifts ($0.8<z<2.0$), observed flux
density at 11 cm greater than 0.15 Jy, and a declination range
$-3^{\circ}<\delta$(1950)$<+60{^\circ}$.  This sample was known as the
``damped \Lya~sample''.

Changes to the original target list and our modes of observing,
however, were required during the execution of the survey in response
to updated information about our originally proposed targets, the
telescope, and the spectrographs.  The events leading to modification
in the list of proposed targets and modes of observation included the
following: changes in object brightnesses (i.e., improved ground-based
photometry indicating a quasar was not as bright as originally
believed, either because of errors in past photometry or variability
of the source); discovery of the spherical aberration due to the
primary mirror, requiring a change in spectrograph aperture in order
to retain spectral resolution at the cost of increased exposure time
per object; lower than anticipated efficiency of the Blue side of the
Faint Object Spectrograph (FOS) initiating an attempt to switch the
far UV spectroscopy observations to the Goddard High Resolution
Spectrograph (GHRS); temporary loss of side 1 (the far UV detector) of
the GHRS, requiring a switch back to the FOS at the cost of increased
exposure time and hence a reduction in the number of targets that
could be observed; installation of COSTAR, providing a favorable
change in spectrograph aperture and reduction of observing overhead,
but bringing a decrease in the total efficiency of the combined
observing system; changes in the list of targets protected by
guaranteed time observers (GTOs) as they adapted their own programs;
cycle-to-cycle variations in the amount of observing time recommend by
the TACs and assigned by STScI to the survey. The cumulative effect of
these modifications was the nearly complete elimination of the
secondary ``damped \Lya~sample'' of targets from the survey, a
decrease in the number of objects in the ``primary sample'', and a
reduction in the number of far UV observations that could be made. The
latter observations were particularly expensive in observing time, but
were also the only means of probing the lowest redshifts ($z<0.3$) for
\Lya~absorbers.  A total of 76 quasars was finally observed as part of
the executed Key Project observations.

In addition to the Key Project, John Bahcall led several GTO programs
designed to investigate many of the problems of interest to the Key
Project.  These observations used identical observing modes and were
planned by a subset of the Key Project team. A result of the close 
coordination in the design of the two programs was the ability to
plan complementary observations, including modifications of the GTO
program to complete observations that had to be dropped from the
original Key Project plan of observations.  Twenty-two objects were
observed as part of the Bahcall GTO program, 19 of which met the Key
Project's criteria for inclusion in the originally defined ``primary
sample''. Six of these objects were partially observed under the Key
Project and had additional observations made as part of the GTO
program.

The merged data from the GTO and Key Project surveys resulted in the
final sample of 92 observed objects listed in Table~1.  We have
included the three GTO objects that have redshifts lower than the
original sample definition criteria because the data that were
obtained are in all other respects comparable with the typical data in
the survey.  In Table~1 we present basic information about the 92
quasars that are investigated in this study. We include the
V\'eron-Cetty \& V\'eron (1991) catalogued redshifts (with the
exception of PG~1407$+$265 and \pg2302~as detailed in the notes on
individual objects (\S\ref{sec-notes}); also note that since 1991
improved redshifts have become available for some of the objects in
the sample, but the redshifts in Table~1 are those that were used in
our analysis), $V$-band magnitudes, and B1950 coordinates.  In
addition we have listed the objects' Galactic coordinates and their
J2000 coordinates which were provided by the Space Telescope Science
Institute (STScI). The coordinates were measured from the digitized
version of the ``Quick~$V$'' Survey plates (Epoch 1982) described in
Lasker \etal~(1990) using STScI's Guide Star Selection System
Astrometric Support Package (GASP) and should be accurate to better than
1$''$.  Additional information listed in Table~1 includes the UT date
on which each object was observed with a particular grating (indicating
the wavelength region observed, see next section) and the observed
continuum flux at a representative wavelength covered by that
observation. If a strong emission line is present at the listed
wavelength a measurement was made at the adjacent continuum with this
shift noted in a footnote to Table~1.  If there is not an entry
under a particular grating no observation was made in that mode. In
Figure~1 we display, on an Aitoff projection, the Galactic coordinates
of all 92 objects observed as part of the quasar absorption line
survey.

\subsection{Observing Modes}

All observations in the survey were made using the FOS of the
\HST. The observing procedures are described in detail in Paper~II. A
description of the FOS is given by Ford \& Hartig (1990).  The FOS has
two Digicon detectors, denoted as ``Blue'' and ``Red'', which were used
for observations below and above 1600~\AA, respectively (with one
exception being the use of the Blue detector for the G190H observation
of PKS~0405$-$123).  We used three higher resolution gratings (G130H,
G190H, and G270H, $R~=1300$) to observe between 1150--3270~\AA\ and a
low resolution grating (G160L, $R \approx 180$) to observe some
objects between 1150--2400~\AA.  Prior to the December 1993
\HST~servicing mission, all observations were made using the
$0.25^{\prime\prime}~\times~2.0^{\prime\prime}$ slit aperture
(effectively $0.25^{\prime\prime}~\times~1.4^{\prime\prime}$ in size
as the height of the diodes limited the length of the slit). The
resulting full width at half maximum (FWHM) of unresolved lines
observed with the different gratings are the following: 1.1~\AA\
(G130H), 1.5~\AA\ (G190H), 2.0~\AA\ (G270H), and 9.4~\AA\ (G160L).
After the servicing mission (cycle~4 observations), we changed to the
$0.3''$ circular aperture (despite the designated name for the
aperture, the size when used with COSTAR was $0.26''$) resulting in 
changes in the FWHM of unresolved lines to the following: 0.9~\AA\ (G130H),
1.4~\AA\ (G190H) and 1.9~\AA\ (G270H). No observations were made for
our program using the G160L grating in cycle~4.

\subsection{Calibration of the Spectra}
\label{sec-calibration}

The details of the calibration of the spectra are discussed in
\papertwo, but we review a few of the critical steps in this section:
corrections for noisy diodes, the setting of the wavelength zero-point,
and corrections for small scale (and time dependent) variations in the
sensitivity of the detectors.

Occasionally some observations were affected by the presence of a
``noisy'' diode. Once identified, such diodes were generally disabled
and would not affect future observations. However, if a diode
intermittently malfunctioned during an observation, one could usually
mitigate the effects during processing of the data: since a long
integration was divided into sub-units and the effect of a noisy diode
were easy to identify, it was generally possible to exclude the
affected subsets of the data and combine the rest.  Corrections of
this sort were made to the following spectra: NAB~0024$+$22 (G270H);
PKS~0637$-$75 (G190H); US~1867 (G190H, G270H); PG~0953$+$415 (G130H);
TON~28 (G130H); 1130$+$106Y (G160L); PG~1202$+$281 (G190H);
PG~1216$+$069 (G130H); PKS~1252$+$11 (G160L); PG~1634$+$706 (G190H);
PKS~2300$-$68 (G270H); and PKS~2344$+$09 (G190H).

We have placed all the higher-resolution observations for a given
object obtained with the different gratings (\eg~G190H and G270H
spectra) on a common wavelength scale by requiring that the strong,
singly ionized interstellar medium (ISM) absorption lines are at rest
(zero redshift; see Paper II for details).  In Table~2 we list for
each spectrum the wavelength zero-point-offsets that were {\it added}
to the reduced spectra.  No correction could be made for some of our
observations (noted in Table~2) because the necessary Galactic
absorption lines were not detected or well measured. In the final
column of Table~2 we list the small additional offset that should be
added to our spectra and line lists to place them in the heliocentric
rest frame. These additional offsets are based on work described in
Savage \etal~(1993; hereafter Paper~III) and in Lockman \& Savage
(1995).  Vacuum wavelengths are quoted throughout this paper.

Corrections for variations in sensitivity of the photo-cathodes and
diodes of the FOS detectors are made from flat-field exposures
obtained from observations of stars.  Because the variations in
sensitivity were found to be time dependent, we attempted to use the
flat-field data generated from calibration observations that used the
same spectrograph aperture (since the corrections are also aperture
dependent) and obtained closest in time to the quasar observations.
However, there were periods when the interval between calibration
observations was too long to enable us fully to correct some portions
of some spectra.  The variations in sensitivity are particularly
strong in the central portions of the G190H spectra. Residual errors
from the flat-fielding process dominate the statistical noise in some
portions of the spectra and can create spurious weak features (cf.
Paper~II and Jannuzi \& Hartig 1994).  Features that are sufficiently
strong to be in the ``complete sample'' of lines (defined below), but
are recognized by comparison with observations of standard stars and
other quasars to be residual flat-field features (cf. Jannuzi \&
Hartig 1994) are listed at the end of the line lists for each object
and are indicated by ``FF'' in both Table~3 and Figure~2.

The calibrated data consist of two arrays: the flux and the 1$\sigma$
uncertainty in the flux as a function of wavelength.  Figure~2 shows
the fully calibrated \HST~data used to construct the database
presented in this paper. From each spectrum and its associated flux
uncertainty array an equivalent width detection limit array is
constructed (see \S\ref{sec-select} and Paper~II for the definition
and further discussion).  This array for each object is also shown in
Figure~2.

\section{Constructing the Third Catalogue of Absorption Lines}
\label{sec-constructing}

In this section we review the methods used to select, measure, and
identify the absorption lines that comprise our catalogue.

\subsection{Selection and Measurement of the Absorption Lines}
\label{sec-select}

Throughout our selection and measurement of the absorption lines in
our spectra we have used well-defined and extensively tested
algorithms developed by Key Project scientists in order to minimize
the subjective elements in the absorption line measurement and
identification processes.  The details of our line selection and
measurement software are described in \papertwo~and in \S3 of CAT2.
The notation for the line-fitting algorithms in this paper is the same
as that used in Paper~II.  The summary presented in \S3 of CAT2 is an
accurate description of our procedures, and we follow that discussion
closely in this review of the more important aspects of our
procedures.  The few minor changes to the analysis software since
CAT2, all introduced because of the complexity of some of the spectra
of high-redshift quasars, are detailed in this subsection.

The fitting of the continuum remains by far the most subjective aspect
of our entire analysis pipeline.  This is particularly true in the
vicinity of strong emission or absorption lines. The continuum fits
for the objects presented in this third catalogue were in general
determined by at least three members of the team whose alternative
fits were merged into a consensus fit by SK and BTJ.  The adopted
continuum fit for each object is shown as a dotted line in Figure~2.
In addition to defining the continuum, the fitted curves attempt to
reproduce the emission line profiles.

A normalized spectrum for each of the quasars was created by dividing
the calibrated data by the continuum fit.  The fluxes in unresolved
lines (absorption or emission) in the normalized spectrum were
calculated using the procedure described in \S6 of Paper~II.  The
Spectral Spread Functions (SSFs) in all of the Key Project
observations have Gaussian profiles (see Figure~1 of Paper~II).  Each
line is assigned a significance level, $SL$:
\begin{equation}
SL~=~{\vert W \vert \over {\overline \sigma}(W)}~~~, 
\label{eq:siglevel}
\end{equation}
where $W$ is the observed equivalent width, and ${\overline
\sigma}(W)$ is the 1$\sigma$~error in the observed equivalent width of
an unresolved line.  Note that ${\overline \sigma}(W)$ is calculated
with the flux errors at the positions of strong absorption lines
replaced by the errors interpolated from the surrounding continuum
points (see Paper~II).  The significance level differs from the more
familiar definition of signal-to-noise ratio, $SNR = | W | /
\sigma(W)$, because of the use of the {\it interpolated} error array
for the calculation of~$SL$.  This search procedure is iterated a
number of times (nine times in the analysis described in this paper)
so that features consisting of several closely spaced lines can be
separated into individual components.

At this point one constructs the function $\sigma_{\rm
det}(\lambda_{\rm obs})$ (the 1$\sigma$ detection limit for an
unresolved line) from the array of~1$\sigma$~uncertainties in the flux
(${\overline \sigma}(W)$; see Figures~9 and~10 in Paper~II).  A
preliminary line list is created using the features whose equivalent
widths exceed a selected $SL$ threshold
\hbox{($W > C_{SL} \sigma_{\rm det}(\lambda_{\rm obs})$};
$C_{SL}$~=~3.0 for the preliminary line lists in this paper).  Each of
these lines is fitted with a variable-width Gaussian profile; this
allows one to characterize the properties of resolved lines.  Note
that this procedure was designed to detect and characterize
unsaturated lines (see Paper~II) and that parameters for individual
lines whose profiles significantly deviate from the SSF or Gaussian
shape cannot be reliably measured with this technique.

Some minor modifications had to be made to the version of the line
measurement code used in the generation of the line lists presented in
this paper because of the complexity of the spectra of the high
redshift objects.  Occasionally the Gaussian fitting software used for
CAT2 would completely fail to decompose complicated blends into
individual components.  The Gaussian fitting procedure attempts to fit
a blend with as few lines as possible to produce a reasonable
representation of the data; a blend is fitted first with one line (the
strongest component in the initial SSF search; see Paper~II), then
successive components are added until the fit does not significantly
improve.  When some complex features were modeled with one component,
the fit was so poor that the algorithm did not converge; this of
course produced a rejection of the line.  The software has been
modified so that if the fitting algorithm does not converge for any
single SSF line, the blend is fit using the two strongest components
as starting parameters, and the additional components are added in
successive iterations just as in our previous studies.

We refer to the collection of lines having $SL~>~4.5$ in a given spectrum 
as the ``complete sample.''  The minimum observed
equivalent width as a function of wavelength that a line must have to
be included in the complete sample is:
\begin{equation}
W_{\rm min}({\rm complete~sample}; \lambda ) ~=~ 4.5 \times
\sigma_{\rm det}(\lambda )~~~,
\label{eq:MinCompleteSample}
\end{equation}
i.e., $W~>~W_{\min}(\lambda)$.  We have also constructed lists of
lines with $3.0~<~SL~<~4.5$; these unpublished lists are referred to
as ``incomplete samples''.  We do not use the lines in the incomplete
samples in our statistical analyses, although we occasionally remark
on such lines in the notes on individual spectra.  In Figure~2 we
display for each source the adopted $4.5\sigma_{\rm det}$ minimum
equivalent width limit as a function of wavelength that must be
exceeded in order for an absorption line to be included in the
complete sample (cf.  Paper~II).

If the Gaussian fitting software identified a feature with only a
single component, we counted that feature as a single line, even
though the FWHM of such lines are occasionally unphysically large.  At
higher resolution, we expect these broad features to break up into
many components.  Since blended, broad features are more likely to
occur at the higher line densities found at the higher redshifts (and
a much larger fraction of the objects in this third catalogue are at
$z>1.0$, compared with CAT1 and CAT2) we may be slightly under-counting
the $z \sim 1.3$ systems relative to lower redshift systems.

When an absorption line was detected in more than one spectrum (i.e.,
in both the G190H and G270H spectra), only the more accurate measurement
(in general this was the G190H observation) is presented in Table~3.

Our combined catalogue (CAT1, CAT2, and this paper) now includes
measured line lists for 78 of the 83 quasars observed with the higher
resolution gratings. The only objects not yet included are five of the
six BAL quasars (PG~0043$+$039 is in CAT1) whose continua and broad
absorption lines have proved difficult for our automated software
to analyze.

\subsection{Lyman$-$Limit Systems Search and Measurement Software}
\label{sec-limitmeasure}

The spectra presented in CAT1 were examined for the presence of
Lyman$-$limit systems (LLSs) with an automated software algorithm (see
Paper~II for details).  In total ten LLSs were identified, and an
analysis combining the HST results with previously published {\it
International Ultraviolet Explorer} (IUE) and ground-based LLSs
observations was presented in Stengler-Larrea \etal~(1995; Paper~V).

We have made two minor modifications to the Key Project LLS search
procedure since the publication of CAT1.  The redshift is now
calculated using an ``effective'' LLS rest wavelength of 914~\AA\
instead of the 912~\AA\ used in CAT1; the new value attempts to
compensate for the blending of the high-order lines in the Lyman
series that occurs at the FOS resolution.  This, of course, leads to a
slight reduction of the measured redshift.  We have also modified the
rest wavelengths used in the flux-ratio calculation; in CAT1 the
``lower'' and ``upper'' bands were 842--902~\AA\ and~922-982~\AA,
whereas for this paper we adopted 883--903~\AA\ and~933--953~\AA. The
widths of the bands were narrowed because examination of the FOS data
showed that this produced measurements that were more acceptable to
the ``eye'' than the earlier technique (especially on the
short-wavelength side), and the short-wavelength cutoff on the upper
band was moved redward to avoid the continuum roll-over produced by
blending of the high-order Lyman lines.  The major change from the
CAT1 results is that the measured optical depths are, on average,
slightly higher; this is primarily due to the lowering of the flux in
the short-wavelength band.

All of the survey spectra were searched for LLSs using the new
parameters; the results are discussed in \S\ref{sec-limit}.

\subsection{Identification of Absorption Lines}
\label{sec-identify}

The procedures for identifying the absorption lines for CAT3 remain
essentially the same as those described in \S4 of CAT2.  Our
procedure is embodied in the software package called {\it ZSEARCH},
described in CAT1, Bahcall \etal~(1992a), and CAT2.  Our goal is to
create a set of physically sound identification algorithms that can be
applied objectively and efficiently to analyze large numbers of
simulated spectra that have the same characteristics as the observed
data.  We only outline here the main aspects of the procedure we use
and highlight improvements or required modifications that have been
made since CAT2 was published.  Many of these improvements are
required by the higher absorption line density in the spectra
considered in this paper relative to the spectra in CAT1 or even CAT2.

The standard ultraviolet absorption lines that we considered as
possible absorbers are the strongest allowed, one-electron, dipole
transitions from ground or excited fine-structure states of cosmically
abundant elements.  Since the number of accidental coincidences
increases with the number of standard lines considered, we included
only standard lines that are likely to have significant equivalent
widths in our spectra.  We used the standard line search list
presented in Table~7 of CAT1 with the modifications listed in CAT2 and
the addition of higher Lyman series lines (\Lyz, \Lyeta, \Lyth, \Lyi,
\Lyk, and \Lyl) that were not included in the CAT2 version of
\zsearch.

To be identified as absorption due to a heavy element (in this paper
we will refer to such lines as metal lines) or as members of the
Lyman series of hydrogen, lines must have observed wavelengths that
are consistent with the identification and are required to have
relative equivalent widths consistent with their known $f$-values and
the uncertainties in the line measurements (CAT2).  The maximum
allowed discrepancy in wavelength is either 1~\AA\ or $3
\sigma(\lambda)$, whichever is larger (where $\sigma(\lambda)$ is the
root-mean-square wavelength measurement error in the line center, see \S6 of
\papertwo~for a complete definition; note that any uncertainty in the
wavelength zero-point-offsets is not included in this term).

The identification of absorption lines consists of four phases: 

1.) {\it ZSEARCH}~identifies candidate Galactic interstellar lines, which
can constitute a significant ``background'' within which lines of
extragalactic origin must be recognized.  

2.) The version of {\it ZSEARCH}~used for CAT2 considered every line
in the spectrum that was not identified with a Galactic ISM line to be
a candidate \Lya~line at the appropriate redshift, $z_{\rm cand}$, and
checked for lines that might be associated metal or Lyman series
lines.  Note that $z_{\rm cand}$ is constrained not to exceed the
emission-line redshift of the quasar by more than 10,000~\kms.  The
CAT2 version of {\it ZSEARCH}~would not allow an extragalactic
identification to supersede one as a Galactic ISM line.  For CAT3 we
modified this step to test {\it all lines} as possible \Lya~lines,
including those tentatively identified as Galactic ISM lines.  This
allowed for the possibility that an extragalactic absorption line
might not only be blended with a Galactic ISM line, but actually be
the dominant source of absorption at that wavelength.  This was found
to be necessary for {\it ZSEARCH}~to be able to identify correctly the
absorption line systems in the spectra of the sources with emission
redshifts larger than approximately 1.3.  Under the rules of CAT2 some
metal line systems which require the presence of \Lya~to be identified
would have been rejected because their \Lya~line had already been
identified as Galactic ISM absorption.

3.) The third pass through the line list by {\it ZSEARCH}~is used to
perform independent searches for the strongest expected metallic
doublets (C~IV, N~V, O~VI, Mg~II, Al~III, Si~IV, and Zn~II) that are
not found in the second phase.  This allows for the identification of
metal line systems for which the \Lya~line is not observable.

4.) Finally, {\it ZSEARCH}~makes a fourth pass through the line list,
testing individually all matches of standard lines with observed lines
for candidate associated-absorption systems, i.e., for systems with
redshifts within $3000 $ \kms of the emission-line redshift.

Despite the many strengths of {\it ZSEARCH}, we have not yet been able
to construct complete identifications for the lines in eight of the
higher redshift (all have $z>1.0$ and 6 have $z>1.6$) quasars in our
sample: PG~0935$+$416, MARK~132, TON~34, Q~1101-264, PG~1206$+$459,
S4~1435$+$63, PG~1715$+$535, and PG~1718$+$481.  The spectra of these
objects share one or more of the following properties that complicate
their analysis: some have complex continua, sometimes the result of
heavily blended lines, that are difficult to fit; the high-redshift
objects ($z>1.6$) can have \Lyb~lines in the observed spectrum that
are indistinguishable from \Lya~lines because the entire \Lya~path to
the emission redshift of the quasar was not observed; many of the
higher redshift objects contain LLSs (including six of the eight for
which our identifications are incomplete), further reducing the
observed path length in each spectrum and adding numerous metal lines
to the spectra so increasing the blending with \Lya~lines.  The low
spectral resolution of our observations exacerbated all of the
above problems.  There is a ninth object that shares many
characteristics with the eight just listed.  The lines in the spectrum
of UM~18 ($z=1.89$) have a similar degree of blending and multiple
possible identifications. In spite of these difficulties, we were able
to generate a complete set of identifications for the lines in this
spectrum. However, we caution that the identifications for this object
are significantly less robust than those for the other objects in the
catalogue.  Efforts are in progress to improve our methods in order to
obtain more complete identifications of the lines in these spectra.

The numerical redshifts given in this paper are calculated as in CAT2,
by iterating, for multiple-line systems, the redshift initially found
by {\it ZSEARCH}. The software finds the redshift that minimizes
$\chi^2$, with $\chi^2$~calculated using the unweighted differences of
the observed and predicted wavelengths. The redshift with the
minimum $\chi^2$ is adopted and a final set of identifications based
upon this redshift is determined. In CAT1, no iteration was performed
to determine the best-fit redshift for multiple-line systems.

The probability for accidentally identifying metal-line absorption
systems that satisfy all of the self-consistency rules is generally
small.  Even if there are only three lines in a candidate redshift
system (\eg~\Lya~plus a strong doublet) the probability of these three
lines being a chance identification is typically much less than
5\%~(see Table~4 and 5 in CAT2; \S3.4, Table~4, and Table~5 in this
work).  Isolated doublets (doublet systems for which additional lines
are not present in the line list) that occur at wavelengths shortward
of the quasar \Lya~emission can have a significant probability of
being the product of two \Lya~absorbers and we discuss such ``false
identifications'' in \S3.4.  Furthermore, in some of the more complex
cases of line identification considered in CAT3, multiple candidate
metal line systems have measured features in common.  While the
arguments used in CAT2 and \S3.4 below still justify the belief that a
candidate system with 12 lines is real and not a chance coincidence,
the membership of each individual line in that real system is less
certain, particularly when some lines can plausibly be associated with
more than two secure (i.e., multiple lined) redshift systems.  If one
line appears to occur in more than one absorption system with
comparable plausibility, then we tabulate the dominant identification
with the standard line and redshift but indicate in the notes on
individual objects (\S4) and/or Table~3 that the line is blended.  It
is in resolving these multiple identification issues among several
plausible identifications that errors in the identifications of
individual lines will undoubtedly be made.

Absorption lines that lie shortward of the \Lya~emission line are
generally presumed to be \Lya~or higher members of the Lyman series
unless they are identified as Galactic interstellar lines or as metal
lines in a multiple-line absorption system.  Candidate C~IV doublet
systems without additional lines in the system were occasionally
identified, but the significant probability that these lines are
actually \Lya~lines is noted with a ``p'', indicating that there is a
specific probability (see \S3.4 for discussion) that the lines are are
due to \Lya~systems. Similarly, pairs of tabulated~\Lya~lines that
could alternatively be identified as part of a C~IV doublet are
indicated with a ``p''.  Finally, a small number of lines shortward of
the \Lya~emission line were left unidentified when there was some
secondary evidence that the line might not be real (\eg~an isolated
line with a high $SL$, but a low SNR; such lines are often the result
of slight errors in the continuum fit found only after the continuum
fitting stage of the analysis has been completed). In such cases the
identification was usually left blank.  An identification that is part
of an absorption system with multiple lines is preferred
algorithmically over an identification with only one other line.

Examples of how the identification rules were applied in practice can
be found in the notes on individual spectra that are presented in
(\S\ref{sec-notes}). The identifications of the lines are listed in
Table~3.

\subsection{Reliability of C~IV and O~VI Identifications}
\label{sec-reliability}

We have performed Monte Carlo simulations with pseudo-C~IV or -O~VI
doublets to estimate the probability that a pair of absorption lines
might accidentally have the appropriate properties to be identified
either as C~IV or
an O~VI absorption doublet.  The technique used is
identical to that described in \S5~of CAT2 with the one change that
instead of 500 simulations per tested spectrum and type of simulation,
1,000 were performed.  The procedure described in CAT2 preserves all
of the complexities in the observed spectra and in the identification
algorithms while allowing us to estimate the probability of accidental
identifications with pseudo-doublets.

An individual simulation to test for false doublet identifications
consists of identifying the lines in an object's complete sample of
observed lines with the same software used for the normal
identifications (\zsearch), but with a modified list of standard
wavelengths for the line search list.  Specifically, the doublet whose
identification statistics we wish to test is removed from the standard
line list and is replaced with a pseudo-doublet. The properties of
each pseudo-doublet are a rest frame separation that ranges randomly
between that of the tested doublet (\eg~2.57~\AA~in the case of C~IV)
and 50~\AA, and the ratio of the oscillator strengths of the two lines
is reversed relative to the doublet being tested.

We searched for pseudo-doublets in all of the line lists of the
observed spectra presented in CAT1 and this paper (simulations for the
four objects in CAT2 are presented in that paper) with the same
algorithmic software that was used to make the line identifications
described in \S\ref{sec-identify}.  Searches were performed on line
lists purged of multiple line absorption systems (metal line systems,
systems with multiple Lyman series lines, and Galactic ISM lines).  As
discussed in CAT2, the multiple line systems have a negligible
probability of being false identifications (i.e., chance systems) and
their lines must be removed from the list prior to running the
simulations as these lines would not be mistaken for C~IV or O~VI (or
any other doublet). Leaving them in the line list to be tested would
result in a large over-estimate of the number of expected chance
systems.  The lines which remain are those that might be identified
with real doublets or pseudo-doublets; they are potential \Lya~lines
(without \Lyb) and miscellaneous unidentified lines longward of the
\Lya~emission line.  Note that candidate C~IV (with or without
associated \Lya~absorption) and O~VI doublets that do not have
additional metal lines associated with them are left in the line list
that will be tested as it is possible that these might not be secure
identifications.

The generated numbers of chance pseudo-doublets can be used as an
estimate of the number of false C~IV and O~VI doublet identifications
in our final line lists.  The pseudo-doublets and any incorrectly
identified C~IV and O~VI doublets are being produced predominantly by
the chance distribution of \Lya~lines. Any clustering among the
\Lya~absorbers on scales larger than the natural splitting of the
doublet might produce a slight increase in the number of
pseudo-doublets detected.  While there is evidence for some clumping
of the \Lya~absorbers around extensive metal line systems (over
velocity scales of a few thousand kilometers per second; CAT2, Jannuzi
1998), we expect this effect to be small.

The results of the simulations provide a guide to the reliability of
our identifications of C~IV and O~VI doublets. We first describe the
results for C~IV doublets and then address the less complicated case
of the O~VI doublets. For the C~IV doublets we are interested in two
types of false identifications: 1) C~IV doublet plus \Lya~absorption,
2) C~IV doublet for which any associated \Lya~absorption is not
accessible. Multiple line systems (more than three lines) including
the C~IV doublet have a negligible chance of being false systems and
are not considered in this discussion.

Two sets of 1,000 simulations for pseudo-C~IV doublets were created
for each object.  The first set was designed to determine the
probabilities of finding false C~IV doublets with \Lya~absorption and
to understand how many C~IV doublets would be identified if the
requirement that \Lya~be present when observable were dropped.  This
set of simulations is identical in technique to those done in
CAT2. The mean number of systems that would contain both lines of the
pseudo-C~IV doublet$+$\Lya~is one of the values returned. This is
generally a very small number, as it is rare that three lines randomly
have the correct properties. The number of chance systems with the
C~IV doublet redward of the quasar \Lya~emission was also determined
by this set of simulations, and was in all cases negligible. As was
done for CAT2, the simulation software also determined the average
number of pseudo-doublets per simulation that were found when we
dropped the requirement that \Lya~be present when accessible. Normally
our identification software tests a candidate C~IV doublet by
requiring that if associated \Lya~absorption would be observable
(appropriate wavelength coverage provided by the spectrum and of high
enough SNR that a \Lya~line as strong as the stronger of the C~IV
lines would be detected) it is detected. If it is not possible to test
for associated \Lya, then a candidate C~IV doublet without any
additional lines in the system could be accepted.  Note that the mean
number of systems found without making this test is an upper limit on
the mean number of expected false systems that would be found by our
identification software.

The test for the presence of related \Lya~is not generally or
uniformally made in the literature. The numbers of chance doublets
found (see Table~4) is relatively large and shows that the test should
always be applied if appropriate data are available.

To reproduce more closely what occurs when we make our identifications
a second set of simulations for pseudo-C~IV was run using a smaller
input line list limited to lines shortward of the quasar
Lyman-$\alpha$ emission for which, if identified as C~IV, any
associated \Lya~would not be accessible.  This population of candidate
C~IV doublets is the largest source of misidentified C~IV
doublets in the combined catalogue.

In Table~4 we show the results of simulations for the incidence of
chance C~IV systems; the two lines for each object refer to the two
types of simulations.  The columns in Table~4 are as follows: (1) the
name of the quasar; (2) the quasar's emission redshift; (3) the number
of potential lines that might be identified with pseudo-doublets; (4)
the mean number of systems per spectrum that were found to have
pseudo-C~IV and an associated \Lya~absorber; by construction this was
zero for the second set of simulations for each object; (5) the
average number of pseudo-C~IV identifications per simulation that were
found when we dropped the requirement that \Lya~be present when
observable (first line) and the mean number of pseudo-C~IV doublets
that would be expected in the spectrum without any associated
\Lya~line being found (second line); (6)-(12) the percentage of the
total number of simulations in which 0-6 individual pseudo-C~IV
identifications were found (without \Lya); (13) the mean expected
number of chance or false identifications of C~IV doublets in the
given spectrum, the result of summing the mean number of a chance C~IV
doublets {\it with} an associated \Lya~absorber and the mean number of
C~IV systems that would have been accepted without \Lya~absorption
because the expected associated \Lya~absorption was not observable
(due to lack of wavelength coverage in the spectrum or poor SNR in
that region of the spectrum). 

An example will help clarify the contents of Table~4. For 3C~57 there
are 23 lines in the line list purged of multiple line systems. Four
additional lines are dropped when making the test for pseudo-C~IV
doublets for which checking for \Lya~absorption would not be possible
because any \Lya~line associated with a candidate pseudo-doublet
containing one of these lines would be observable.  The first set of
simulations yield a mean number of 0.01 pseudo-C~IV doublets with an
associated \Lya~absorber detected. In addition an average of
1.09 pseudo-doublets was found when the test for associated
\Lya~absorption were ignored.  The second set of simulations for 3C~57
reveal that a slightly lower number of false C~IV doublets without
\Lya~identifications is actually expected (0.97) when the rules are
applied in exactly the same manner used when generating the real
identifications.  Thirty-six percent of the simulations yielded no
pseudo-doublet systems without \Lya~absorption, but 20.2\% of the
simulations contained two pseudo-C~IV doublets.

There is a range in accidental probability of an order of magnitude
depending upon which spectrum is being searched for pseudo-doublets.
The average number of pseudo-C~IV doublets found in the real
(observed) spectra varies from 0.0 to 2.35 (for PG~1634$+$706,
$z=1.334$, simulations in CAT2) per spectrum within the region that
includes absorption by \Lya~absorbers and is negligible outside of
this range.  The numbers of pseudo-C~IV doublets found in the observed
spectra, and hence the implied number of false C~IV identifications,
are sufficiently large that for some applications of the catalogue
they must be taken into account in the analysis.

In the combined catalogue there are 21 C~IV doublets for which no
other associated lines are identified and for which a check for
associated \Lya~absorption could not be made (due to lack of
wavelength coverage or SNR in the spectrum).  Similar to these 21 C~IV
doublets there are also pairs of \Lya~lines listed in the tables of
identified lines in CAT1, CAT2, and CAT3 (Table~3, this paper) that
could have been identified as being part of a total of 25 C~IV
doublets.  The choice between being identified as a C~IV doublet
without any other associated lines or as \Lya~absorption was not
rigorous and we have selected the identification that seems most
likely for each case (based on the goodness of fit to the C~IV doublet
and the plausibility of the absence of other metal lines), but noted
that formally an alternative identification would meet all of our
identification rules.  In CAT3 Table~3 such systems are listed as
either ``C~IV,p'' (candidate C~IV, associated \Lya~not observable,
might be \Lya~lines) or ``\Lya,p'' (candidate \Lya, possibly C~IV
doublet).  The ``p'' indicates that there is a specific probability
that the identification of the indicated line should be \Lya~(or
C~IV). The results in Table~4 can be used to determine this specific
probability, and we make use of this in our paper on the evolution of
\Lya~absorbers (Weymann~\etal~1998).  In addition to the systems
listed in Table~3, there are similar systems among the objects
considered in CAT1 and CAT2. In general these systems in CAT1 and CAT2
are not indicated in the line identification tables with a special
note or symbol, although four of the five pairs of lines that occur in
PG~1634$+$706 and one of the three pairs in PG~1352$+$011 are noted in
the CAT2 tables.  For completeness, we now list the objects from CAT1
and CAT2 that have such candidate C~IV or pairs of \Lya~lines and the
wavelengths of the lines comprising such systems: PKS~0044$+$03
(1928.38, 1931.94~\AA), 3C~454.3 (1786.37, 1789.08~\AA), TON~153
(1995.65, 1999.03, 2029.16, 2032.06,  2034.49, 2036.95, 2038.76~\AA), PG~1352$+$011 (1846.54,
1848.97, 1851.97, 1894.26, 1896.77~\AA), PG~1634$+$706 (2412.04,
2416.80, 2453.03, 2457.65, 2557.44, 2560.88, 2564.86, 2599.88,
2603.34, 2755.05, 2759.96~\AA).

For the purposes of this section of the paper only, we will consider
all of the C~IV doublets without other lines that have a significant
probability of actually being two \Lya~lines (the ``C~IV,p'' systems)
and the pairs of \Lya~lines that could have been identified as C~IV
doublets (the ``\Lya,p'' lines) as possible false identifications as
C~IV doublets.  There are a total of 46 doublets of this type in the
completely identified line lists in the combined catalogue.

We can now consider the reliability of the C~IV doublet
identifications in our catalogue and estimate the fraction of the
identified C~IV systems that might be the result of the chance
matching of \Lya~or other lines.  There are 107~tabulated C~IV
absorption line systems in the combined catalogue of absorption
systems, but since our simulations can only be generated for
completely identified line lists we must restrict ourselves to 70
objects.  There are 97 tabulated C~IV systems in this subset of the
entire catalogue and an additional 25 ``\Lya,p'' systems, for a total of
122 candidate C~IV systems.  Fifty-nine of these systems occur redward
(longward) of the quasar's \Lya~emission and 63 occur at wavelengths
blueward (shortward) of the quasar \Lya~emission.  We will consider
the reliability of these identifications as a function of being
blueward or redward of the quasar redshift.

Of the 63 C~IV systems observed blueward of the quasar \Lya~emission,
14 have an extremely small probability of being chance
identifications:  nine systems which include both lines from the doublet,
\Lya~absorption, and additional associated metal line absorption
and/or higher Lyman series lines; three systems have both lines from
the doublet plus multiple other metal lines detected; two systems have
one line of the C~IV doublet detected (the stronger, C~IV~\l1548.20),
\Lya~absorption, and additional metal lines.  Three additional systems
(composed of both lines from the C~IV doublet, an associated
\Lya~absorption line, but no additional lines) have a finite
probability of being a chance system (in the entire survey, on
average, only 0.48 systems of this type are expected, including both
blueward or redward systems). The remaining 46 systems have a larger
chance of being false identifications. These systems include 21 ``C~IV,p''
and 25 ``\Lya,p'' systems.  Based on the results of our simulations we
estimate that in the spectra of the 70 quasars with completely
identified line lists we would expect 20.97$\pm4.6$ of the 46
candidate C~IV systems to be the result of the chance alignment of
other lines.  In summary, approximately one third of the C~IV doublets
which occur blueward of the quasars' \Lya~emission are expected to be
\Lya~lines.

Understanding the chance number of systems is more straightforward for
the 59 systems with their C~IV doublets redward of \Lya~emission.
Among the 59 are 32 systems with the C~IV doublet, \Lya, and
additional metal and/or H~I (Lyman series) absorption lines detected.
The probability for a chance system with these characteristics is
negligible. Similarly, the broad system in the spectrum of
PG~2302$+$029 is certainly a real system (see the next section and
Jannuzi~\etal~1996). There are 11 additional C~IV doublets observed
redward of the quasar \Lya~emission for which \Lya~absorption is also
detected, but no other metal lines are associated with the system. For
these systems the simulations reveal that there is close to zero
probability that there is a false identification. For two of the 59
redward C~IV doublets any associated \Lya~absorption was not
accessible in our observed wavelength region. Based on our
simulations, we expect 0.5 such systems to occur in our sample drawn
from the 70 completely identified quasar spectra. Finally, the 59
redward C~IV systems include 13 identifications that have only the
strong line of the C~IV doublet detected. Twelve of these include
absorption by \Lya; a subset of four systems include additional metal
lines. In the 13th system the expected wavelength of associated
\Lya~absorption was not observable, but the system does have the
weaker line of the C~IV doublet in our incomplete sample. We estimate
for the observed singlet redward C~IV systems and those redward C~IV
doublet systems that do not have detected associated \Lya~absorption
that a total of 2.5$\pm$1.0 systems are the product of a chance
occurrence of other absorption features.

Combining the results of the simulations for false C~IV
identifications, we estimate that among the 122 C~IV systems
(including the 25 \Lya,p pairs) in the completely identified
line lists there are 24$\pm$5~that are false identifications.

A similar analysis leads to an estimate of the number of false O~VI
identifications included in our sample of 31 O~VI systems (from a
total sample of 41) contained in the spectra of the 70 objects for
which identifications are complete. The simulations of pseudo-O~VI
systems were carried out as described in CAT2 and the results are
presented in Table~5.  For pseudo-O~VI doublets, the average number
per spectrum varies from 0.0 to 0.08.  Except for the case of UM~18
(see below), all of the identified O~VI pseudo-doublets are associated
with detected \hbox{Ly-$\alpha$}~lines, as is required by our identification
rules.  Relatively few accidental O~VI doublets are found in the
observed spectra, principally because of the limited redshift range in
which the O~VI doublets are observable and because \Lya~is nearly
always accessible when O~VI is detected and provides a strong 
constraint against chance identifications.  For all but one of the
objects for which we performed simulations any candidate O~VI doublet
would have to have its associated \Lya~accessible. However, for the
highest redshift quasars in our sample, and specifically for UM~18,
this is not the case for the entire O~VI path length; at the highest
redshifts any \Lya~associated with the candidate O~VI system moves
beyond the red edge of the G270H~observations. This explains why the
mean number of chance O~VI systems without \Lya~is different from the
mean number with \Lya~only for UM~18; for the rest of the higher
redshift quasars the line identifications are not complete and we did
not perform any simulations for these line lists.

Twenty-two of the 31 identified O~VI systems have both lines of the
O~VI doublet detected and the vast majority of these systems (18) are
extensive metal line systems that have a negligible probability of
being a chance absorption complex. Although the reality of the
associated O~VI is less certain, these systems are probably real
(\eg~see Table~5 for examples of the low probabilities for false
isolated doublets in general). There are 8 systems for which only the
stronger line of the O~VI doublet is in the complete sample, but 6 of
these are also part of extensive metal line systems. While the
association of O~VI absorption with these systems when traced by a
single line is less secure, nearly all of these identified O~VI
systems are expected to be real.  Using our simulations we find that
the sum over all 70 objects of the mean expected number of false O~VI
doublet systems that would contain both lines of the doublet plus
associated \Lya~absorption is only 0.55. 

We estimate that of the 31 O~VI systems in the sample of 70 objects
with completely identified lines, one is a chance coincidence or false
identification.

\section{Notes on Individual Spectra}
\label{sec-notes}

In this section we present descriptions and notes on the spectra
obtained as part of the survey. The notes are grouped by object and
presented in order of the right ascension (B1950) of the object.
Entries are presented for all of the 92 objects, but complete notes,
table of identifications (Table~3), and figures (Figure~2), are only
included for the objects being discussed for the first time or being
reanalyzed because of the availability of additional wavelength coverage
(a new G130H spectrum of PKS~2251$+$11 analyzed together with the G190H
and G270H spectra from CAT1) or to use the same analysis software on
the spectrum (PKS~0405$-$123 from the Bahcall GTO program).

For a small fraction of the detected features in the line lists we
have included comments in \S\ref{sec-notes} that the feature detected
by the line measurement software is not believed to be a real feature,
but rather an artifact of an imperfect fit to the continuum. In the
vast majority of cases these features are broad and shallow, are
probably due to a systematic error in the continuum fit that placed
the fit, and have a SNR in their measured equivalent width that is
less than 4.5, while still having $SL>4.5$. Such slight systematic
shifts of the continuum were not always caught during the review of
the continuum fits, but once the line identification process was
started for a particular line list we did not allow revision of the
continuum fit in order to avoid introducing any additional bias into
the line measurement process.

In general any detected \Lya~absorption due to gas associated with our
Galaxy was not properly measured by the software and in many cases the
depression or dip in the spectrum was ``taken out'' by fitting the
feature during the construction of the continuum fit. When the
Galactic \Lya~absorption was present it was always a damped feature
and contaminated by geocoronal \Lya~emission, and would not have been
well measured by our software in any case. Fitting the feature as part
of the continuum fitting process allowed weaker, nearby intergalactic
\Lya~absorbers to be detected and measured by the software.

{\bf \boldmath UM~18 (RA: 00:02:46.3  DEC: 05:07:29.0, B1950;
$z_{\rm{em}}$=1.89,\unboldmath ~108~lines)} 
\nobreak
Our spectrum of UM~18 is an example of the spectra, all of moderate to
high redshift quasars, that presented the greatest difficulties in
the measurement and identification of their absorption lines.  These
spectra share the properties of heavy line blending, difficult
continua to fit, and limited total observed wavelength coverage. We
will use our discussion of the identification of the lines in UM~18 to
illustrate these points.

The higher redshift objects have a larger density of observed
absorption lines for at least three reasons: 1.) the rest
equivalent width detection limits in the higher redshift quasar
spectra tend to be lower (i.e., more sensitive to intrinsically weak
\Lya~absorption lines) than those of the low redshift quasar spectra
because the survey was designed to deliver spectra of uniform
signal-to-noise ratio (i.e., uniform in observed equivalent width
limit; see Figure~1 of Weymann \etal~1998); 2.) many of the higher
redshift quasars have at least one Lyman$-$limit system in their
spectra, resulting in numerous associated metal lines and higher order
Lyman lines in their spectra; 3.) the higher order Lyman series lines
(\Lyb, \Lyg, etc.) are more readily observable in these spectra due to
a larger observed path length at the higher redshifts (i.e., more of
the less numerous stronger systems are included) and the more
sensitive rest equivalent width detection threshold. All of these
effects are evident in the spectrum of even the moderate redshift
quasar PG~1538$+$477 ($z_{\rm em}=0.770$) as shown in Figure~3 of
Jannuzi (1998).

The higher density of lines combined with the relatively low spectral
resolution of our observations results in severe blending of features,
complicating every aspect of the analysis: the fitting of the
continuum, the selection and measurement of the absorption lines, and
identification of the lines.  For the objects with Lyman$-$limit
systems (including UM~18 and six of the eight objects for which we
have incomplete identifications) the observed path length is also
reduced, limiting our ability to cross check candidate metal line
identifications by looking for supporting lines elsewhere in the
spectrum. Similarly, for any of the quasars with emission line
redshifts greater than about 1.6, our G270H spectrum does not extend
far enough in the red to reach the quasar's \Lya~emission line.  This
results in a region of each of these observed spectra that might
contain \Lyb~lines that are indistinguishable from \Lya~because we can
not check for an associated \Lya~absorber longward of 3200~\AA.

Given the problems discussed above, while we have listed
identifications for the lines in the spectrum of UM~18, we consider
these identifications to be less robust than the complete
identifications presented for most of the objects in the combined
catalogue. As an example, while the strong lines at 2797.57 and
2802.84~\AA~are close to the wavelengths of Galactic ISM Mg~II, {\it
ZSEARCH}~formally rejected this identification in favor of either
\Lya~absorption systems or as a C~IV doublet at $z=0.8071$. There were
also several additional possible identifications for each line.

As we were completing this paper, a G190H spectrum of UM~18, obtained
as part of a program of A. Koratkar and collaborators, became
available in the {\it HST} archive. This spectrum, taken in
spectropolarimetry mode, was reduced in a manner consistent with our
own observations, although since the Al~II~\l1670.79 Galactic ISM
line was not detected we could not check the zero-point of the
wavelength scale.  We used the G190H spectrum to identify the redshift
and measure the properties of the Lyman$-$limit system ($z=0.86$)
listed for this object in Table~6.  This LLS might be associated with
a candidate metal line system present at $z=0.8519$ or one of the
other systems identified between $z=0.85$ and 0.87.  Unfortunately,
the generally poor signal-to-noise ratio of the G190H spectrum yielded
only one line that would normally meet our criteria for inclusion in
the complete sample (at 2011.75$\pm0.76$~\AA, $W=1.54\pm0.56$,
FWHM$=4.24$, $SL=6.5$) if the spectrum had been taken in a manner
consistent with our other observations; this single additional line
does not significantly aid the identifications of the lines in the G270H
spectrum.

The remainder of our notes on the UM~18 spectrum consist of a listing
of alternative identifications for some of the observed lines.  The
discussion of these identifications is not exhaustive, but
serves to demonstrate some of the complications faced when trying to
identify all of the lines in the spectra of the higher redshift
objects.

As we mentioned above, because our G270H spectrum ends near 3200~\AA,
there is a section of the UM~18 spectrum for which unidentified lines
might be \Lyb~lines associated with \Lya~lines between 3200~and
3513~\AA~(the wavelength of \Lya~emission for the quasar). This
wavelength range includes the measured lines between 2760~and
2965~\AA.  While an isolated \Lyb~line in this region could not be
identified without additional data, if additional higher Lyman lines
or metal lines are present such systems could be verified.  This is
the case for the system at $z=1.8011$: we did not observe the region
containing \Lya~absorption, but the system could be identified by the
presence of the \Lyb, \Lyg, \Lyd, and \Lye~absorption lines.  An
analysis including ground-based observations of the near UV portion of
the spectrum would help resolve some of the uncertainties in the
identifications of the lines between 2760~and~2965~\AA.

The lines at 2234.94 and 2238.33~\AA~could also be a C~IV doublet at
$z=0.4433$.  The line at 2334.08~\AA~might be Si~III in a system at
$z=0.8519$ including candidate lines from \Lya, C~II~\l1334.53
(alternatively \Lya~at $z=1.0335$), Al~II~\l1670.79 (observed at
3093.95; alternatively \Lya~at $z=1.5452$), Fe~II~\l1608.45 (more
probably \Lya~at $z=1.4458$ or part of a C~IV doublet at $z=0.9204$),
and Si~IV~\l1393.76 (more probably \Lyb~at $z=1.5160$). This candidate
system would also be in rough agreement with the redshift of a Mg~II
system detected by Churchill, Steidel, \& Vogt (1996), although this
fact was not used in choosing the above identifications and became
known to us after completion of the identifications.  The
2267.12~\AA~line would alternatively be \Lya~at $z=0.8649$. The
\Lya~line at $z=0.8670$ (observed at 2269.66~\AA) has candidate O~I
and Al~II lines at respectively 2431.04 and 3118.56~\AA. The \Lya~line
at 2320.50~\AA~is unusually broad and is likely a blend of several
lines.  The line at 2360.65~\AA~could be \Lyb~at $z=1.3011$~or
\Lya~at $z=0.9419$.  The line at 2383.60~\AA~contains a component from
the Galactic ISM Fe~II~\l2382.77~line, but is a broad blend that
certainly contains additional lines.

The system at $z=1.4721$ might also include the O~VI doublet at
observed wavelengths 2550.59 and 2564.25~\AA, although alternative
identifications were marginally preferred. The \Lyg~line of this
system is alternatively \Lyb~at $z=1.3436$.

The lines at 2418.33 and 2431.04~\AA~might be an O~VI doublet
associated with other lines at $z=1.3436$ (including the \Lya~line
currently listed at $z=1.3438$, which is quite broad and probably a
blend). The broad line (blend?) at 2540.53~\AA~might include N~II at
$z=1.3436$.  The line at 2944.45~\AA~had multiple equally probable
identifications.  The lines at 2973.27 and 2978.78~\AA~might be a C~IV
doublet at $z=0.9204$.  The candidate $z=1.4377$ system currently
includes identifications of absorption from Si~II~\l1260.42 and
Si~II~\l1193.28 at respectively 3072.46 and 2908.15~\AA, but these
lines might alternatively be caused by \Lya~absorption. The metal line
identifications were slightly preferred due to the presence of the
higher supporting hydrogen absorption lines at the same redshift.  The
\Lyg~line (at 2465.91~\AA) from the $z=1.5359$ system is blended with
the Si~II line from the $z=0.9564$ system. The \Lyb~line from the
same system would be blended with the ISM Fe~II line at 2600.89~\AA.
The system at $z=1.500$ might include the O~VI blended with the observed
lines at 2580.59 and 2595.98~\AA. Any associated C~IV in this system
would be redward of the observed spectrum.  The line at 2607.45~\AA~
might include a component of Galactic ISM Mn~II.

{\bf \boldmath PKS~0003$+$15 ($z_{\rm{em}}$=0.450, \unboldmath ~58 lines)}
\nobreak
In total there are 21 \hbox{Ly-$\alpha$}~lines identified along this line of
sight, two of which are associated with detected metal line systems
($z=0.3660$~and 0.4014).  The strong \hbox{Ly-$\alpha$} line in the
$z=0.3660$~system (at 1660~\AA) is likely to have an associated C~IV
doublet contained in the very broad (most likely heavily blended)
absorption complex between 2113 and 2118~~\AA. The expected
wavelengths for the associated C~IV would be 2114.82 and 2118.33~\AA.
The observed \hbox{Ly-$\alpha$} line is intrinsically broader than 270
\hbox{km~s$^{-1}$}; it would not be too surprising for the matching C~IV to
be contained in more than one component.  This heavy blend of
candidate C~IV absorption was fitted by the software as two narrow and
one broad components, but a fit to two doublets would be justifiable.
We identify this complex collectively as C~IV associated with the
$z=0.3660$~system.  In the incomplete sample the \hbox{Ly-$\epsilon$}~line
of this system is present at $1280.90\pm0.21$~\AA~
($W=0.31\pm0.10$~\AA, $SL=4.23$, FWHM$=1.36$~\AA).  The two lines at
1408.87 and 1416.41~\AA~and identified as \hbox{Ly-$\alpha$} lines might
contain O~VI associated with the $z=0.3660$ system.  The region of the
spectrum between the lines at 2179.46 and 2183.15~\AA~was difficult to
fit because of a narrow apparent emission line (probably not real) at
2184~\AA~and the resulting uncertainty in the continuum fit. These
latter two absorption lines might be a C~IV doublet at $z=0.4078$.

Galactic ISM lines in the incomplete sample include Si~III
($1206.37\pm0.23$~\AA; $W=0.69\pm0.23$~\AA; $SL=4.04$;
FWHM$=1.39$~\AA), Si~IV ($1393.82\pm0.20$~\AA; $W=0.19\pm0.06$~\AA;
$SL=3.28$; FWHM$=1.10$~\AA), and Mg~I ($2026.46\pm0.30$~\AA;
$W=0.18\pm0.06$~\AA; $SL=3.68$; FWHM$=1.72$~\AA).  The unidentified
line at 2800.97~\AA~might be Mg~II absorption by gas in a high
velocity cloud associated with our Galaxy. In this, the asymmetry of
the 2803~\AA~line is mirrored in the corresponding 2796~\AA~line,
although it was not fitted separately by the software.  There is
strong \Lya~absorption from the Galactic ISM, but the line does not
appear in the line list because the absorption was ``taken out'' by
the continuum fit.

The average equivalent width of the eight unidentified lines is
$0.23$~\AA, considerably less than the average strength of the
identified lines and significantly less than the average \hbox{Ly-$\alpha$}
line identified in this spectrum (0.63~\AA).  A number of these
unidentified lines may be unrecognized flat-field residuals or the
results of errors in the continuum fit.

{\bf \boldmath NAB~0024$+$22 ($z_{\rm{em}}$=1.118,\unboldmath ~46 lines)}
\nobreak
This spectrum contains 16 identified \Lya~lines; five of these are
associated with detected metal lines.  There is a high excitation
metal line system at $z=1.1102$, approximately 2,300 \hbox{km~s$^{-1}$} from
the redshift of the quasar.  This system contains C~III, the C~IV and
O~VI doublets, as well as \hbox{Ly-$\alpha$}, \hbox{Ly-$\beta$}, 
and \hbox{Ly-$\gamma$}.
Lower excitation lines (\eg~of Si~II, S~II, N II, C II) were not
detected.  This system provides one of the few cases in our data set
where a search for absorption by Ne~VIII~\ll770.4,780.3 is possible; no
statistically significant features are evident at 1625.7 and
1646.4~\AA. However, the detection limits are large (more than 1~\AA)
because of the low signal-to-noise ratio of the spectrum near the
short wave-length edge of the G190H spectrum.  The redshift of this
system is less than 3,000 \kms from the quasar redshift and would meet
most definitions of being an associated absorber.

A metal line system at $z=0.8196$ includes lines from C~III, Si~III,
the strong line of the C~IV doublet, \hbox{Ly-$\alpha$}, \hbox{Ly-$\beta$},
\hbox{Ly-$\gamma$}, and \hbox{Ly-$\delta$}.  There is a candidate C~IV doublet at
$z=0.4069$ associated with the \hbox{Ly-$\alpha$} line at 1710~\AA. The
stronger line of the doublet would be blended with one of the O~VI
lines in the $z=1.1102$ system.

The incomplete sample of lines contains three candidate Galactic ISM
lines.  These are from Al~II~\l1670.79 at
$1670.81\pm0.28$~\AA~($W=0.50\pm0.15$~\AA, $SL=3.22$, 
FWHM$=1.51$~\AA), Mn~II at $2594.26\pm0.36$~\AA~ ($W=0.24\pm0.07$~\AA,
$SL=3.36$, FWHM$=2.04$~\AA), and Mg~I at
$2852.98\pm0.33$~\AA~($W=0.38\pm0.09$~\AA, $SL=3.75$,
FWHM$=2.04$~\AA).

The line at 2419~\AA~is quite broad. If this feature is real it is
likely to be a blend of the listed identification and some other
unidentified feature.

Three of the lines in the spectrum are broad or resolved and have
multiple possible identifications.  Although these lines are likely to
be blends, we have listed only one identification for each in the
table.  The following are possible additional contributors to the
observed absorption: 1685.88~\AA, \hbox{Ly-$\beta$} at $z=0.6432$;
1727.75~\AA, \hbox{Ly-$\gamma$} at $z=0.7760$; and 1773.50~\AA, Si~II at
$z=0.4069$.

{\bf PG~0043$+$039 (\boldmath$z_{\rm{em}}$=0.384)\unboldmath ~CAT1}
\nobreak
One of six BAL quasars observed during the Key Project and Bahcall GTO
program observations; the narrow absorption line identifications are
presented in CAT1 and the object's BAL properties are discussed by
Turnshek \etal~(1994).

{\bf PKS~0044$+$03 (\boldmath$z_{\rm{em}}$=0.624)\unboldmath ~CAT1}

{\bf \boldmath PG~0117$+$213 ($z_{\rm{em}}$=1.493,\unboldmath ~84 lines)}
\nobreak
There are 33 \Lya~lines in this spectrum, seven of which are
associated with detected metal lines.  This spectrum contains two
extensive metal line systems occurring close together in velocity
space. The systems at $z=1.3389$ (seven identified lines, including
the candidate N~II absorption, see below) and $1.3426$ (fourteen
lines) are of the order of 1,000 \hbox{km~s$^{-1}$} apart.  Both systems
include lines from high-ionization states of O~VI. The O~VI~\l1037.62
line of the $z=1.3389$ system is in the incomplete sample at
$2426.72\pm0.37$~\AA~with an $W=0.24\pm0.07$, a FWHM of 1.97~\AA~and a
$SL=3.21$.

The identification of 2271.87 and 3103.72~\AA~as respectively C~III
and C~II in a $z=1.3256$ system should be considered tentative. This
system has both its \Lya~and \Lyb~lines blended with other lines.
There is also a candidate O~VI doublet associated with this system at
2399.18 and 2412.70~\AA.  The very strong line at 2488.89~\AA~is
identified as \hbox{Ly-$\alpha$} at $z=1.0473$. Surprisingly no other
associated lines were detected in the system, although the expected
wavelengths of lines from Si~II, C~IV, O~I, and Fe~II are all
accessible in the spectrum.  The 2574.74~\AA~line, a very broad
feature, is likely a blend of absorption from several lines. While
identified as \hbox{Ly-$\alpha$}, it likely contains O~VI from the $z=1.4952$
system.  The identification of 2700.63~\AA~is uncertain, possibly
being \hbox{Ly-$\alpha$} at $z=1.2215$, C~II at $z=1.0237$, and/or N~I at
$z=1.2503$. The reality of this feature is also uncertain, as it might
be the result of an error in the continuum fit.  While our
identification rules lead to the 2595.24~\AA~line being identified as
Galactic Mn~II, no other lines of Mn~II were detected, including the
stronger line expected at 2576.89~\AA. Therefore it is possible that
this line is intergalactic and caused by \Lya~at $z=1.1351$.  The
matching line of the C~IV~\l1548.20 at $z=0.9676$ (3047.02~\AA) would be
expected to be blended with the observed line at 3050.44~\AA, which is
identified as O~I in the $z=1.3426$ system.

There are candidate systems at $z=0.3159$ (Al~III doublet observed at
2440.62 and 2451.23 \AA), 0.5766 (Si~II~\l1526.72, C~IV
\ll1548.20,1550.77 doublet, Al~II~\l1670.79, and perhaps
Fe~II~\l1608.45), and 0.9400 (\Lya, Al~II~\l1670.79, and perhaps
Si~II~\l1260.42 and C~II~\l1334.53) which include several of the same
lines. With only the information contained in our data we are unable
to firmly identify all of the possibly shared lines.  However, the
presence of Mg~II absorption systems at $z=0.5764$, 0.7289, 1.0478,
and 1.3251 (Churchill 1997) support our independent decision of
listing the $z=0.5766$ identifications in the table when a choice had
to be made.  Alternative identifications for the lines in this system
that have not already been mentioned include the following:
2407.08~\AA~(\Lya~at $z=0.9800$), 2444.84~\AA~(Si~II at $z=0.9400$),
2535.60~\AA~(N~II~\l1083.99 at $z=1.3389$), and 2634.62~\AA~(\Lya~at
$z=1.1672$).  If the $z=0.9400$ system is not real, then its C~II line
would most likely be \Lya~at $z=1.130$ and the Al~II line would not
have an identification.

Several pairs of lines in this spectrum are heavily blended or are
asymmetric lines that our fitting software split into two components.
As a result, equivalent-width measurements of the individual
components are uncertain, although the total equivalent width of the
blended feature is accurately measured. These measured pairs include
the following lines: 2382.92, 2384.98; 2593.13, 2595.24; 2782.44,
2784.87; 2825.69, 2827.49~\AA.

{\bf PG~0122$-$00 (\boldmath$z_{\rm{em}}$=1.070)\unboldmath ~ CAT2}

{\bf \boldmath 3C~57 (RA: 01:59:30.4 DEC: $-$11:47:00, B1950;
$z_{\rm{em}}$=0.670,\unboldmath ~35 lines)}
\nobreak
There are 19 lines identified as \Lya~absorbers in this spectrum; none
are associated with detected metal line systems.  The line at
1934.94~\AA, identified as \Lya, is part of a candidate C~IV system at
$z=0.2498$ that includes this line and the line at 1938.24~\AA. The
lines at 1762.32 and 1766.24~\AA~might be a C~IV doublet at
$z=0.1386$. The line at 1892.42~\AA~is probably the result of an error
in the continuum fit.  The strong and broad feature at
1609.20~\AA~needs to be confirmed as a real feature since it occurs at
the end of the observed spectrum and is subject to a large systematic
error in the continuum fit. If the current measured line properties
were confirmed, the line would be identified as \Lya~at $z=0.3237$
with lines of O~I and C~II observed respectively at 1723.17 and
1766.24~\AA. The line at 1912.64~\AA~is probably an artifact of the
continuum fit.

{\bf \boldmath PKS~0232$-$04 ($z_{\rm{em}}$=1.434,\unboldmath ~49 lines)}
\nobreak
There are 25 \Lya~lines in this spectrum and metal lines are
associated with two of these lines.  The broad and shallow line at
2775.90~\AA~is likely to be a composite of several other lines that
could not be separated at the resolution of our spectrum.  The line at
2686.94~\AA~is blended with other lines, and might not actually be an
independent feature.  The \Lya~lines at 2890.85~and 2925.64~\AA, which
occur in the \Lya~emission line, could be artifacts of the continuum
fit.  If the candidate C~IV system at $z=0.7391$ is not C~IV, then the
line at 2692.23~\AA~would be \Lya~at $z=1.2146$ with a matching \Lyb~
at 2270.76~\AA.  The 2697.12~\AA~line would be \Lya~at $z=1.2186$.
The pair of lines at 2953.80 and 2961.95~\AA~were alternatively
identified by \zsearch~as a candidate doublet of Mg~II absorption at
$z=0.0564$.  \zsearch~provides possible identifications of the lines
at 2976.26~\AA~and 3175.91~\AA~as respectively N~V ($z=1.4021$) and
Si~IV ($z=1.2788$), but in each case the weaker line of the doublet is
not present.  No damped systems or LLSs were detected in the G160L
spectrum of this object.

{\bf 3C~95  (\boldmath$z_{\rm{em}}$=0.614)\unboldmath ~ CAT1}

{\bf \boldmath PKS~0405$-$12 ($z_{\rm{em}}$=0.574,\unboldmath ~46 lines)}
\nobreak
We have rereduced and reanalyzed the FOS spectra of PKS~0405$-$12
previously published by Bahcall \etal~$\;$(1993b) using the current
versions of the line measurement and identification software.  The
most important cause of change in the line list is the manner in which
the complete sample was determined (see \S3.1 and Paper~II for
discussion of the difference between using the ratio of the measured
equivalent width to the uncertainty in that measurement and the
definition we have chosen for the quantity $SL$). Other modifications
included an improved adjustment of the zero point wavelength scale
that resulted in a slight change of the measured velocities of the
extragalactic absorption systems.  A revised continuum fit for the
G270H grating data caused significant changes to the measured strength
of some lines.

Significant differences between the old and new line lists are
the following:

1.) The line at 1408~\AA~is now identified as Si~III at a redshift
of 0.1671, instead of \hbox{Ly-$\alpha$}.

2.) The complex blend at 1252.27~\AA~is now identified as \hbox{Ly-$\alpha$}
at a redshift of 0.0301. Although S~II~\l1253.79~probably contributes
to this blend, the wavelength difference between the expected and
observed position of the line is too great for this line to be
produced entirely by Galactic ISM S~II.  If the identification of this
feature as \Lya~is correct (and it would be valuable to confirm this
with additional observations), then this system might be of particular
interest for detailed study given its proximity.

3.) We have identified three additional metal lines associated with
the $z=0.1671$ system noted by Bahcall~\etal~(1993b) and associated
with two galaxies by Spinrad~\etal~(1993). The system is now known to
include absorption from \Lya, C~IV, Fe~II, Mg~II, Si~II, and Si~III.

4.) Eleven lines appear in the new complete sample that were not in
the Bahcall et al.$\;$ (1993b) line list. Ten of these lines have
equivalent widths less than 0.3~\AA. All of the lines appear in the
new line list because of either the change in the way we determined
the significance of the line or a revision of the continuum fit.  Of
the eleven new lines, two are metal lines associated with the 0.1671
system, one is a Galactic ISM line, five are identified as \Lya~, and
three are unidentified.  The unidentified lines are probably produced
by slight errors in the continuum fit. The reason the continuum fit
errors occur preferentially in the redward portion of the spectrum is
because this region has a higher signal-to-noise ratio and the quasar
continuum emission is not well fitted by a low-order spline.  The
resulting ``false'' lines, that would normally be too weak to be
included in the complete sample have significance levels that are
sufficient to require inclusion in the complete sample.

5.) Six lines in the Bahcall et al.$\;$ (1993b) sample are not in the
new complete sample. Five of these lines are in the incomplete sample
of lines: two retain their identification as Galactic ISM
C~II*~\l1335.71 and Al~II~\l1670.79; two were previously unidentified
lines (at 2711 and 3034~\AA); one was previously identified as a
\Lya~line at $z=0.3633$.  One line from the old sample (at
2605~\AA~and previously identified as Galactic ISM Mn~II) dropped
below a $SL$~of three, and hence out of the incomplete sample, due to
the revised continuum fit. In summary, of these six lines five are
probably real although they no longer appear in the complete sample.

Additional notes on the current identifications follow.  There are 19
lines identified as \Lya~with one of these lines associated with
identified metal lines.  The $z=0.4056$ system might have an
associated Mg~II doublet detected in ground-based data (see Spinrad
et~al.$\;$~1993).  Our revised measurements of the \hbox{Ly-$\alpha$} line
that previously matched in velocity a galaxy in Spinrad
\etal~($z=0.361$) yield a larger separation in velocity than our
original measurement.  The line at 1577~\AA~is formally identified as
\hbox{Ly-$\beta$} at $z=0.5385$, but the line center is more than $3\sigma$
away from the expected position, and this discrepancy might be due to
another \hbox{Ly-$\alpha$} line at $z=0.2977$.  The Galactic ISM
C~II*~\l1335.71 line is in the incomplete sample at
$1335.60\pm0.18$~\AA~($W=0.28\pm0.07$~\AA, $SL=4.00$,
FWHM$=1.10$~\AA).  The Al~II~\l1670.79 Galactic ISM line is also in
the incomplete sample at $1670.83\pm0.21$~\AA~($W=0.30\pm0.06$~\AA,
$SL=4.39$, FWHM$=1.51$~\AA).  The line at 1471~\AA~is identified as
Si~II~\l1260.42 at $z=0.1671$, but the observed width of this line
suggests that it is probably produced by a blend of Si~II with some
other unidentified feature.  The Galactic ISM Fe~II~\l2586.65 line is
affected by a flat-field residual.

{\bf 3C~110 (RA: 04:14:49.2,  DEC: \boldmath$-$06:01:04.0, 1950;
$z_{\rm{em}}$=0.773,\unboldmath ~33  lines)}
\nobreak
 There are 22 \Lya~lines in this quasar's spectrum, but no extensive
metal line systems were detected.

{\bf \boldmath PKS~0439$-$433 ($z_{\rm{em}}$=0.593,\unboldmath ~37 lines)}
\nobreak
This spectrum contains an extensive metal line system at
$z=0.1010$. This mixed ionization system is two to four times stronger
than the absorption found along typical paths halfway through the
Galactic disk/halo gas observed from the position of the Sun.  A total
of 11 lines from this system are in the complete sample.  Two
additional candidate lines of Fe~II (\ll1608.45,2374.46) are detected
in the incomplete sample (at $1770.40\pm0.28$~\AA,
$W=0.29\pm0.09$~\AA, $SL=3.26$, FWHM$=1.51$~\AA; and at
$2614.49\pm0.49$~\AA, $W=0.29\pm0.10$~\AA, $SL=3.94$,
FWHM$=2.47$~\AA).  There are seven Galactic ISM lines in the complete
sample and one line in the incomplete sample (Mg~I, at
$2852.79\pm0.29$~\AA, $W=0.36\pm0.08$~\AA, $SL=3.40$,
FWHM$=2.03$~\AA).  The lines observed at 2865.09 and 3088~\AA~are
blended with respectively the Fe~II~\l2600 line and the Mg~II~\l2803 line
of the $z=0.1010$ system.  They are very likely components of the
Fe~II and Mg~II absorption from this system.  There are seven
identified \hbox{Ly-$\alpha$} lines in this spectrum with two of these lines
associated with identified metal lines.  The C~IV~\l1550.77 of the
redshift 0.4075 system is in the incomplete sample (at
$2182.76\pm0.23$~\AA, $W=0.20\pm0.06$~\AA, $SL=4.26$,
FWHM$=1.63$~\AA).  The C~IV~\l1550.77 of the redshift 0.4268 system is
in the incomplete sample (at $2212.56\pm0.26$~\AA,
$W=0.15\pm0.04$~\AA, $SL=3.41$, FWHM$=1.50$~\AA).

{\bf \boldmath HS~0624$+$6907 ($z_{\rm{em}}$=0.370,\unboldmath ~34 lines)}
\nobreak
There are eleven lines identified as \Lya~lines in this spectrum,
including several at very low redshifts.  The line at 2808.10~\AA~is
on the wing of the Galactic ISM Mg~II~\l2803 line, and is likely just
a result of the Gaussian fit to the line not properly including all of
the absorption in the wings.  The line identified as the Galactic ISM
C~IV~\l1548.20 line might be a blend of both features that were not
well measured in this spectrum. The Galactic ISM Si~IV doublet is in
the incomplete sample at 1393.67$\pm$0.17 and
1403.21$\pm0.20$~\AA~with equivalent widths of 0.23$\pm0.06$ and 
0.18$\pm0.05$ and $SLs$ of 3.85 and 3.28, respectively. The Galactic
ISM Zn~II~(Mg~I) blend at 2026~\AA~is also in the incomplete sample at
2026.56$\pm0.20$ \AA~with $W=0.24\pm0.06$~\AA~and a $SL=4.06$.  The
line at 1808.08~\AA, identified as Si~II \l1808.01, is somewhat stronger
than expected.

{\bf PKS 0637$-$75 (\boldmath $z_{\rm{em}}$=0.654,\unboldmath ~22 lines)}
\nobreak
Nine \Lya~lines are identified in this spectrum, one of which is
associated with the single extragalactic metal line system detected
along this line of sight. This heavy element system has a redshift of
0.4168 and has the strong line of the C~IV doublet, \hbox{Ly-$\alpha$}, and
Si~III in the complete sample.  The incomplete sample contains the
C~IV~\l1550.77 line observed at 2197.18 \AA~with an equivalent width of
$0.24\pm0.06$~\AA~and a $SL=4.1$. Note that the Si~III line at
$z=0.4168$ is blended with the \hbox{Ly-$\alpha$} line observed at 1711~\AA;
this affects its measured properties.

In addition to the Galactic ISM lines in the complete sample
(including lines from Al~II, Fe~II, and Mg~II), the incomplete sample
contains lines from the Zn~II (Mg~I) blend at 2026~\AA~(observed at
2025.42~\AA, $W=0.18\pm0.056$ \AA) and Mn~II~\l2576.88 (observed at
2576.15~\AA, $W=0.35\pm0.08$ \AA).  The presence of high velocity gas
along this line of sight ($l=286.4^{\circ}$, $b=-27.2^{\circ}$) in the
velocity range between $+230$ to $+270$ \hbox{km~s$^{-1}$} has previously
been reported by (Bajaja \etal~1985).  We therefore expect to see the
combined absorption from local gas and this high-velocity gas
extending from $-40$ to $+270$ \hbox{km~s$^{-1}$}. This is reflected in our
observations of the Galactic ISM lines in several ways.  First,
Al~II~\l1670.79 and four of the Fe~II lines have very broad
profiles. Second, the Fe~II~\l2600.17 line and Mg~II lines are resolved
into two components.  Additional discussion of the Galactic ISM lines
can be found in Savage \etal~(1998).

{\bf \boldmath B2~0742$+$31 ($z_{\rm{em}}$=0.462,\unboldmath ~12 lines)}
\nobreak
The Galactic ISM Mg~II absorption lines occur on top of a strong
quasar emission line. Together with the non-Gaussian shapes of these
absorption lines, the difficulties in fitting the continuum are probably
responsible for the line fitting software splitting the very strong
and broad Mg~II~\l2796 line into two lines, one at 2796~\AA~and
the second at 2790~\AA.  The line at 1600~\AA~occurs on the blue edge
of the wavelength coverage of the G190H grating, a region of
particularly low signal-to-noise ratio in almost all our G190H spectra
and in this quasar's spectrum a particularly difficult region in which
to fit the continuum.  The feature at 2025~\AA, identified as being
produced by Galactic ISM Zn~II, is quite broad and unlikely to be due
exclusively (if at all) to absorption by Zn~II.

{\bf \boldmath PKS~0743$-$67 ($z_{\rm{em}}$=1.51, \unboldmath  ~76 lines)}
\nobreak
There are 44 \Lya~lines identified in this spectrum; four are
associated with detected metal lines and a fifth is less than 300 \kms
from the velocity of a metal line system.  There are at least three
systems that include absorption by the C~IV doublet along this line.
There is a very strong blend near 2435~\AA~that was separated by the
analysis software into a non-unique fit consisting of a broad line and
two narrower lines.  With the current data we are unable to identify
the broad component of the fit at 2432~\AA.  The line at 2290~\AA~
might be a continuum artifact.  The line at 2612~\AA~is heavily
blended with the line at 2615~\AA~and might be a subcomponent of the
2615~\AA~system.  Several additional lines in the complete sample
might be continuum artifacts since slightly different versions of the
continuum fit developed during analysis of this spectrum caused these
lines to disappear.  These lines are at 2464.36, 2734.18, and
2738.69~\AA. These last two lines comprise a candidate C~IV doublet
that lacks additional supporting lines. The Galactic ISM Mg~I~\l2852.96
line is blended with a stronger \Lya~line at $z=1.3473$.

{\bf US~1867 (\boldmath$z_{\rm{em}}$=0.513)\unboldmath ~ CAT1}

{\bf \boldmath OJ~287 (RA: 08:51:57.2, DEC:$+$20:17:58, B1950;
$z_{\rm{em}}$=0.306,\unboldmath ~9 lines)}
\nobreak
No extragalactic absorption lines are identified in the complete
sample. There are eight Galactic lines in the complete sample and one
in the incomplete sample (Al~II~\l1670.79, at $1670.40\pm0.20$~\AA,
$W=0.72\pm0.20$~\AA, $SL=4.37$, FWHM$=1.55$~\AA).  Planned G130H
observations of this object were unsuccessful because this object
became too faint at the time the observations were scheduled. There
are candidate C~IV absorption lines in the incomplete sample.  The
line at 1934.89~\AA~is near a feature in the flat-fields, but is
stronger than the typical residual seen in Key Project data.  Future
STIS observations of this normally bright (but variable) object (it is
a well known BL~Lacertae object) would be valuable.

{\bf \boldmath NGC~2841~UB3 (RA: 09:16:30.0 DEC: $+$51:18:53, B1950;
$z_{\rm{em}}$=0.553,\unboldmath ~26 lines)}
\nobreak
There are twelve \Lya~lines identified in this spectrum, one of which
is resolved and associated with a strong metal line system detected at
$z=0.5116$. Lines from Si~III, C~II, and the C~IV and Si~IV doublets
are detected in the complete sample.  The stronger line of the N~V
doublet is present in the incomplete sample (at 1872.48$\pm0.28$~\AA,
$W=0.14\pm0.04$~\AA, FWHM$=1.51$~\AA, $SL=3.27$). The line at 1639.68~\AA~and
identified as \hbox{Ly-$\alpha$} at $z=0.3488$~might be blended with
N~II~\l1083.99 associated with the $z=0.5116$ system.
Galactic ISM lines are detected from Fe~II, Mg~II, and Al~II.
It is possible that the broad unidentified line at 2596~\AA~is
associated with Galactic ISM Mn~II, but this identification was not
made by \zsearch~because the other expected Mn~II lines were not
present in the line list. However, this region of the spectrum is
blended with the wings of the strong Fe~II Galactic ISM absorption.
Blending plus uncertainties in the continuum fit might cause the other
two lines of the Mn~II triplet to have been missed.

{\bf \boldmath PG~0935$+$416 ($z_{\rm{em}}$=1.937,\unboldmath ~131 lines)}
\nobreak
The continuum fit for this spectrum was particularly difficult, and
given the low spectral resolution of the observations and the high
density of lines, is probably inadequate in several regions, adversely
affecting the measurement, separation, and identification of the
lines. Despite these problems, this spectrum also provides the only
damped \hbox{Ly-$\alpha$} system, $z_{\rm{abs}}=1.3720$, in the entire
survey.  The damped line is not well fitted by our software, which
assumes Gaussian line profiles for all lines.  As a result, an
additional line was included in the software's automated fit, at
2889.27~\AA. The resulting redshift of the system (which includes
absorption from C~II, C~III, N~I, O~I, Fe~II, Si~II, Si~III, \Lya, and
\Lyb) is therefore different from that measured from a proper fit to the
damped absorption, $z=1.396$, which we include in \S5, our general
discussion of damped systems.  In addition to the detected damped
system, there is an extensive metal line system at $z=1.4649$ with an
associated Lyman$-$limit system.  The properties of the LLS as
determined by the LLS fitting software are listed in Table~6.

The numerous possible identifications for many of the lines prevented
us from completing the identifications. Even for the lines we have
identified, many are likely to be blended with absorption from other
species.  For example, three identifications for the line at
2317.90~\AA~are C~III at $z=1.3720$ (our current identification),
Fe~III in a candidate system at $z=1.5085$, and C~IV at $z=0.4946$.
Higher spectral resolution observations, perhaps with the Space
Telescope Imaging Spectrograph (STIS), of this interesting line of
sight would be valuable.

{\bf \boldmath PG~0953$+$415 ($z_{\rm{em}}$=0.239,\unboldmath ~28 lines)}
\nobreak
Galactic interstellar absorption lines are identified from ten
different ions: C~II, C~IV, N~I, O~I, Al~II, Mg~I, Mg~II, Si~II,
Si~III, and Fe~II.  In the incomplete sample, Si~IV~\l1393.76 is
present at $1393.75\pm0.16$~\AA~with an equivalent width of
$0.26\pm0.06$~\AA~and $SL=4.13$.  The line at 1190.37~\AA~includes a
contribution from Si~II, but its large equivalent width and FWHM
suggest that it is possible that the line is a blend of Si~II and
S~III~\l1190.21.  The continuum was particularly difficult to fit in
the region between 2570~\AA~to 2610~\AA, making the measured
equivalent widths of lines in this region more-than-usually uncertain
and possibly affecting how many lines the software required to fit the
region. Among the lines with particularly large uncertainties in their
equivalent widths is a line at 2593~\AA.  There is some indication
that this strong line might be Mn~II~\l2594.50 since there is a line
in the incomplete sample that could be identified as Mn~II~\l2576.88
(this line occurs at 2576.48~\AA, has an equivalent width of
$0.30\pm0.11$~\AA, and $SL=4.49$), but there is no sign in the
spectrum of the expected Mn~II~\l2606.46, so the identification as
Mn~II is formally rejected.  The line identified as Galactic ISM
Mg~I~\l2852.96 is unusually strong and broad.

A total of five \hbox{Ly-$\alpha$} absorption features are identified in the
complete sample, although three of the five are too broad (FWHM
larger than the instrumental resolution) to be single, unresolved
\hbox{Ly-$\alpha$} absorbers.  There is a \hbox{Ly-$\alpha$} absorption line at
$z=0.2336$, only 1300 \kms from the emission line redshift.  The
\hbox{Ly-$\beta$} line for this system appears in the incomplete sample at
$1265.6\pm0.153$~\AA~with an equivalent width of 0.24$\pm0.07$~\AA~
and a significance level of 4.4, very close to the level of 4.5
required for inclusion in the complete sample.

There is weak evidence in some of the other G270H quasar spectra taken
contemporaneously with these data for a flat-field contribution to the
line we identify as \hbox{Ly-$\alpha$} at $z=0.0934$ (at 1329~\AA), but there is
no evidence for a flat-field feature in the standard star observations
at the same epoch.

The unidentified line at 2028~\AA~is part of a blend and falls on top
of a quasar emission line. Both of these occurrences contribute to an
additional unquantified uncertainty in the measured equivalent width.
Despite the wavelength agreement of this line with the expected
position of Al~III~\l1854.72 at a redshift matching that of the
\hbox{Ly-$\alpha$} line at 1329~\AA~($z=0.0934$), this identification is
rejected because of the absence of any of the other metal lines that
might also be expected when Al~III is detected.

{\bf \boldmath 3C~232 (RA:09:55:25.5 DEC:$+$32:38:23, B1950;
$z_{\rm{em}}$=0.533,\unboldmath ~24 lines)}
\nobreak
Burbidge \etal~(1971) first drew attention to the close projection of
3C~232 and the nearby bright Sb~III spiral galaxy NGC 3067.  The
separation of the quasar and the galaxy center is $1.'9$ on the sky,
well beyond the extent of the visible disk.  Haschick and Burke (1975)
discovered narrow $(\leq 5.5$ \hbox{km~s$^{-1}$}) H~I 21 cm absorption at a
redshift near 1420~\hbox{km~s$^{-1}$}, close to the mean redshift of
the galaxy; this was confirmed by Grewing and Mebold (1975), and
followed by higher resolution observations (Wolfe 1979; Rubin,
Thonnard \& Ford 1982).  In high-dispersion optical spectra Boksenberg
and Sargent (1978) found Ca~II H and K absorption at nearly the same
redshift as the H~I but of substantially broader (and structured)
velocity profile.  This was the first instance of metal absorption
lines in a quasar spectrum observed at large extension from an
identified galaxy.  From the large ratio of N(Ca$^+$)/N(Na$^0$)
obtained from their data (inferred from upper limits on the equivalent
widths of Na I) they suggested that the absorption arises either in
gas in the outer halo of the galaxy or in high-velocity gas in the
plane.  Low-dispersion ultraviolet observations of $3{\rm C}\ 232$
made with the IUE satellite (Bergeron, Savage, \& Green 1987) provided
a detection of Mg~II in this system and set limits on the presence of
C~IV and Si~IV.  Further optical observations by Stocke \etal~(1991),
in which both Ca~II and Na~I were detected, resolved the system into
two and possibly three components spread over $\sim 160$ \hbox{km~s$^{-1}$},
and from archival {\it IUE} low-dispersion spectra they added Fe~II to
the list of detected lines.  The morphology of the absorbing gas was
clarified by Carilli, van Gorkom \& Stocke (1989) who obtained a VLA
H~I map showing a remarkably long and disturbed tail extending from
the galaxy and appearing as a chance projection at the quasar position
(Stocke et al.~1991).

This absorption system is strongly detected in our data, and includes
lines from Al~II, Fe~II, Mg~II and Mg~I.  There are five \hbox{Ly-$\alpha$}
lines in the spectrum, one of which is identified with a strong
associated metal line system at $z = 0.5314$ (480 \hbox{km~s$^{-1}$} from the
redshift of the quasar).  The 1242 \AA\ line of the N V doublet in
this system is in the incomplete sample (at $1903.34 \pm 0.22$ \AA, $W
= 0.32 \pm 0.08$ \AA, $SL = 4.13$, FWHM$= 1.51$~\AA), while the
1550.77~\AA\ line of the C~IV doublet is blended with the Galactic ISM
Fe~II~\l2374.46 line.  The \hbox{Ly-$\alpha$} line at $z = 0.5167$ probably
is related to weak Fe II and possibly Mg II absorption also found by
Boksenberg and Sargent (1978) in their spectrum.  The Galactic ISM
Fe~II~\l2260.78 line is in the incomplete sample (at $2261.61 \pm
0.38$~\AA, $W = 0.38\pm 0.11$ \AA, $SL = 3.20$, FWHM$= 2.04$~\AA).  A
GHRS spectrum has been obtained of 3C~232 and is being analyzed by
J. T. Stocke and collaborators.

{\bf \boldmath MARK~132 (RA: 09:58:08.1, DEC:$+$55:09:05.0, B1950;
$z_{\rm{em}}$=1.754,\unboldmath ~36 lines)}
\nobreak
The low signal-to-noise ratio of the spectrum, heavy blending of the
lines, and limited observed wavelength coverage (exacerbated by the
presence of a LLS in this spectrum) made the fitting of the continuum
quite uncertain and the possible identifications for some lines too
numerous to allow complete identification of the line list at this
time.  However, the strong hydrogen lines associated with the LLS
observed at $z=1.7327$, as well as some of the other strong lines in
the spectrum, could be identified.  The \hbox{Ly-$\lambda$} line of the
$z=1.7327$~system is in the incomplete sample at
$2508.65\pm0.27$~\AA, $W=1.03\pm0.20$, FWHM$=1.97$~\AA, and $SL=4.12$.
The blend at 2504.68~\AA~is dominated by a blend of the higher Lyman
series lines.  The \Lyb~line undoubtedly includes a contribution from
Galactic ISM Mg~II~\l2803.53.  C~III from the $z=1.7327$~system is a
possible identification for the line (2669.22~\AA) currently
identified as \Lya~at $z=1.1957$, but it is the only possible metal
line identification associated with the LLS.  The lack of identified
associated metal lines with this system, given its estimated column
density, is atypical of the systems found in our survey.  The
properties of this system as derived from the LLS software are listed
in Table~6.

{\bf \boldmath 0959$+$68W1 ($z_{\rm{em}}$=0.773,\unboldmath ~31 lines)}
\nobreak
This spectrum contains 20 identified \Lya~lines. The line at
1658~\AA~is a composite of several features in a noisy region of the
spectrum. Possible contributors to the feature include
\Lyb~at $z=0.6179$ and a known flat-fielding residual.
The lines at 1771 and 1809~\AA~might both be artifacts of the
continuum fit.

{\bf \boldmath TON~28 (RA: 10:01:12.0, DEC:$+$29:13:00, B1950,
$z_{\rm{em}}$=0.329,\unboldmath ~30 lines)}
\nobreak
There is a total of twelve \hbox{Ly-$\alpha$} lines in this spectrum,
including several of the lowest redshift strong lines in the entire
Key Project sample (at 1220 and 1234~\AA).  These lines are excellent
targets for followup investigations to study in detail the nature of
low redshift \hbox{Ly-$\alpha$} absorbers.  The pairs of lines located at
1474, 1475~\AA~and 1495, 1496~\AA~are heavily blended.  The line
observed at 1548.48~\AA~is predominantly \hbox{Ly-$\alpha$} at a redshift of
0.2738, although there is probably a small amount of Galactic ISM
C~IV~\l1548.20, since the weaker line of this doublet is detected in the
incomplete sample at $1551.20\pm0.20$~\AA~($W=0.13\pm0.04$~\AA,
$SL=3.37$, FWHM$=1.10$~\AA).  Another candidate Galactic ISM line in
the incomplete sample is Si~IV at
$1403.05\pm0.17$~\AA~($W=0.18\pm0.05$~\AA, $SL=3.95$,
FWHM$=1.11$~\AA).  There is strong Galactic \Lya~absorption that was 
not included with the current continuum fit (see figures).  A very
broad and shallow feature near 2338.73~\AA~was not included in the
final line list because it is an artifact of the continuum fit.

{\bf \boldmath 4C~41.21 (RA: 10:07:26.1 DEC: $+$41:47:24, B1950;
$z_{\rm{em}}$=0.613,\unboldmath ~36 lines)}
\nobreak
This line of sight includes 18 \hbox{Ly-$\alpha$} lines, of which two have
associated metal line systems including the C~IV doublet and a third
has an associated Si~II line.  All of the lines between 1630~\AA~and
1665~\AA~were difficult to measure because of the lower signal in this
region and the difficulty of placing the continuum fit.  The lines at
1677, 1820, and 1866~\AA~might be alternatively identified as
respectively N~I, O~I, and C~II at $z=0.3979$.  The line at
1937~\AA~is strongly affected by a flat-field residual, but does match
in wavelength the expected position of Si~IV from the $z=0.3902$
system. There is weaker evidence that the line at 1948~\AA~is
influenced by a flat-field residual, but the evidence is not
sufficient to remove the line.  For all of the metal line systems
identified in this spectrum any associated Mg~II absorption would be
observable in ground-based spectra.  The broad non-Gaussian wing of
Galactic Mg~II~\l2796.35 is producing the feature the line fitting
software found at 2797.56~\AA.

{\bf \boldmath PG~1008$+$133 ($z_{\rm{em}}$=1.287,\unboldmath ~42 lines)}
\nobreak
There are 25 lines identified as \Lya.  It is difficult to determine
the reality of the multiple candidate C~IV systems in this spectrum
because the wavelengths at which associated \Lya~absorption might be
observed are not in the available spectrum and there is a significant
probability that one to two of the identified systems are just the
chance coincidence of \Lya~lines (see \S3.4 and Table~4).  For each of
the observed doublets the alternative identification for the lines
would generally be \Lya.  The Galactic ISM Fe~II~\l2382.77~line is very
strong and may be blended with a \Lya~line.

{\bf \boldmath TON~34 (RA: 10:17:06.0 Dec: $+$27:59:00.0, B1950;
$z_{\rm{em}}$=1.924,\unboldmath ~67 lines)}
\nobreak
The heavy blending of the lines in this spectrum and limited observed
wavelength coverage lead to numerous possible identifications for
almost every line and here prevent complete identification of the line
list.  The \Lyg~line for the $z=1.6420$ system is in the incomplete
sample at 2569.71$\pm0.28$~\AA~with $W=0.68\pm0.14$, FWHM$=1.97$~\AA,
and $SL=4.39$. Several metal line systems were identifiable. We now
consider the candidate LLS listed in CAT1 not to be a real system (see
Paper~V for discussion) and no longer list it as part of our
Catalogue. See Table~6 for a complete listing of LLSs found in the
survey.

{\bf 4C~19.34 (RA: 10:22:01.6, DEC:$+$19:27:35, B1950; \boldmath
$z_{\rm{em}}$=0.828)\unboldmath ~ CAT1}
\nobreak
Note the slight revision (see Table~6) of the properties of the LLS
first reported in CAT1. These are as a result of the changes made to
the software measuring the LLSs and not due to any change in the
reduction of the spectrum.

{\bf \boldmath 4C~06.41 (RA: 10:38:40.9, DEC: $+$06:25:58, B1950;
$z_{\rm{em}}$=1.270,\unboldmath ~39  lines)}
\nobreak
The G270H spectrum contains 22 \Lya~lines.  Given the short observed
path length at the higher resolution, it is difficult to test the
reality of all the candidate C~IV doublets observed in this spectrum
because additional lines from these systems do not fall in the
spectral range covered by our higher resolution observations. We have
identified these systems in the table, but caution that the lines
might alternatively be due to \Lya~systems.  The lines comprising the
candidate C~IV system at $z=0.4415$ are almost certainly real as they
are very likely are associated with the Lyman$-$limit system detected
in the G160L spectrum.  The LLS is measured to have a redshift of
$z=0.44$ (see Table~6 for other measured properties).  The weaker line
of the candidate C~IV doublet at $z=0.4928$ is affected by a known
flat-field feature. If this line is entirely a flat-fielding artifact,
or if this doublet is a chance match of two otherwise unrelated lines,
then the lines at 2310.89 and 2315.40~\AA~would be \Lya~at $z=0.9009$
and 0.9046. The C~IV doublet at $z=0.6093$ would be \Lya~at $z=1.0493$
and 1.0530.  The \Lyb~line of the \Lya~line at $z=1.2452$ might be
blended with the lines near 2301~\AA.

{\bf 3C~245.0 (\boldmath RA: 10:40:06.0, DEC:$+$12:19:15, B1950;
$z_{\rm{em}}$=1.028)\unboldmath ~ CAT1}

{\bf \boldmath PG~1049$-$005 ($z_{\rm{em}}$=0.3570,\unboldmath ~15 lines)}
\nobreak
The C~IV system at $z=0.3414$ has an associated \hbox{Ly-$\alpha$} line in
the incomplete sample at 1631.33~\AA~with an equivalent width of
$0.52\pm0.14$~\AA~and $SL=3.55$.  There are a large number of Galactic
ISM lines detected, including lines from Al~II, Zn~II, Fe~II, Mn~II,
and Mg~II. In the incomplete sample are additional Galactic ISM
Fe~II~\l2249, \l2260, and \l2374 (observed at $2249.44\pm0.33$,
$2260.30\pm0.29$, and $2373.80\pm0.29$~\AA~with equivalent widths of
$0.39\pm0.10$, $0.14\pm0.5$, and $0.44\pm0.10$~\AA~and $SL=3.72$,
3.09, and 4.23) and Mn~II~\l2576.88 (observed at $2576.51\pm0.30$~\AA~with
an equivalent width of $0.30\pm0.07$~\AA~and $SL=4.02$).  The line at
2107.34~\AA~is located on top of a narrow emission line, and might be
structure in the emission line rather than an absorption feature.

{\bf PKS~1055$+$20 (\boldmath$z_{\rm{em}}$=1.11)\unboldmath ~ CAT1}
\nobreak
Note the slight revision (see Table~6) of the properties of the LLS
first reported in CAT1. These are as a result of the changes made to
the software measuring the LLSs and not due to any change in the
reduction of the spectrum.

{\bf \boldmath 3C~249.1 (RA: 11:00:27.4, DEC: $+$77:15:08, B1950;
$z_{\rm{em}}$=0.311,\unboldmath ~48 lines)}
\nobreak
This line of sight contains 15 \hbox{Ly-$\alpha$} lines (one associated with
metal lines) and 19 Galactic ISM lines.  The line at 1402.64~\AA,
although matching the expected wavelength of Galactic ISM Si~IV, is
not identified as Si~IV since the stronger line of the doublet is not
present.  This interpretation is supported by the weakness of
C~IV~\l1548.20 observed at 1548.07~\AA: $W=0.19\pm0.04$~\AA.  Galactic
ISM lines from C~II* (at $1335.80\pm0.20$~\AA, $W=0.18\pm0.05$~\AA,
$SL=3.61$, FWHM$=1.10$~\AA) and Si~II (at $1193.17\pm0.17$~\AA,
$W=0.55\pm0.14$~\AA, $SL=3.97$, FWHM$=1.11$~\AA) are in the incomplete
sample.  The line at 1465~\AA~may be an artifact of the continuum
fit. The line at 1657.18~\AA~was identified by \zsearch~as possibly
the stronger line of a C~IV doublet at $z=0.0706$. However, this
identification was rejected because the supporting \Lya~line would be
blended with the line at 1301~\AA; which was identified as Galactic
ISM O~I.

{\bf \boldmath Q~1101$-$264 ($z_{\rm{em}}$=2.148,\unboldmath ~71 lines)}
\nobreak
The heavy blending of the lines in this spectrum and limited observed
wavelength coverage (exacerbated by the presence of a LLS in this
spectrum) made the fitting of the continuum highly uncertain and
caused some problems for the line fitting software (\eg~near 2800 and
3150~\AA), where alternative fits to the blends are certainly possible
and would be chosen by the software with only slight changes in the
continuum fit.  The numerous possible identifications for almost every
line prevented complete identification of the line list here.
We are working on using supplementary ground-based observations to
assist in the line identifications, but that analysis is beyond the
scope of the current paper.  We do include tentative identifications
of the strong lines at $z=1.8377$ associated with the LLS observed at
$z=1.84$, as well as some of the other strong lines in the spectrum.
Additional properties of the LLS are listed in Table~6. The broad line
at 2802~\AA~includes Galactic ISM Mg~II~\l2803.

{\bf \boldmath MC~1104$+$167 ($z_{\rm{em}}$=0.634,\unboldmath ~23 lines)}
\nobreak
There are nine \hbox{Ly-$\alpha$} systems along this line of sight, one of
which ($z=0.4549$) includes metal lines from C~IV and Si~III. The
feature at 1966~\AA, which occurs on the side of a strong quasar
emission line, is probably an artifact of the continuum fitting
process. If the line is real, it would be \hbox{Ly-$\alpha$} at a redshift of
0.6179.  There is a candidate Galactic ISM Zn~II line in the
incomplete sample at $2062.09\pm0.22$~\AA~($W=0.17\pm0.04$~\AA,
$SL=4.05$, FWHM$=1.51$~\AA)

{\bf \boldmath PG~1116$+$215 ($z_{\rm{em}}$=0.177,\unboldmath ~37 lines)}
\nobreak
The spectrum of Galactic ISM lines detected along this line of sight
is among the richest found in the Key Project sample of
observations. A total of 25 Galactic ISM lines are identified in the
complete sample and Al~III (at $1855.09\pm0.40$~\AA,
$W=0.15\pm0.06$~\AA, $SL=4.03$, FWHM$=2.23$~\AA) and Zn~II (at
$2062.51\pm0.21$~\AA, $W=0.14\pm0.03$~\AA, $SL=4.01$, FWHM$=1.51$~\AA)
are in the incomplete sample.  There is strong Galactic \Lya~
absorption that has an approximate observed equivalent width of
2.96~\AA.

There are seven lines identified as extragalactic \Lya~lines, two of
which have an associated metal line.  The line identification software
accepts the line at 1203~\AA~as the stronger line of the O~VI doublet
associated with the \hbox{Ly-$\alpha$} line at 1417.79~\AA, $z=0.1663$,
although there are no other lines in the complete sample to provide
additional support for this identification.  In the incomplete sample
of lines there is a possible match to the expected positions of the
C~IV doublet, but at very low significance level. The candidate O~VI
line might be part of an associated absorption line system being
produced either by gas intrinsic to the quasar or associated with the
group or cluster of galaxies that is associated with the quasar
(Jannuzi 1997).

There is weak evidence that a flat-field residual might be affecting
the measurement of the line at 1454~\AA, but not enough to remove the
feature from the complete sample. If this line were produced by a
\hbox{Ly-$\alpha$} absorber it would have a 
redshift over 5000 \hbox{km~s$^{-1}$}~
larger than the emission redshift of the quasar.  All of the detected
lines from 1393 to 1454~\AA~occur on top of a strong quasar emission
line and are particularly uncertain due to the subjective nature of
the continuum fit. In particular, the line at 1434.96~\AA~can be made
to appear or disappear completely with slightly different continuum
fits.  The lines at 2386.12 and 2798.74~\AA~are not identified, but
are blended with or near strong Galactic ISM lines and might be
related to high velocity Galactic ISM clouds. Alternatively they might
be a result of our choice of fitting Gaussian profiles to the strong
non-Gaussian profiles of the strong ISM lines, leaving portions of the
wings to be fitted as separate lines.

{\bf UM~425 (RA: 11:20:46.6, DEC:$+$01:54:17, B1950; \boldmath
$z_{\rm{em}}$=1.465)\unboldmath ~ }
\nobreak
This is one of the six BAL quasars observed by the Key Project and
Bahcall GTO program.  The spectrum is displayed in Figure~2, but the
analysis of this spectrum will be presented in Turnshek \etal~1998.

{\bf \boldmath 1130$+$106Y, ($z_{\rm{em}}$=0.51,\unboldmath ~22 lines)}
\nobreak
Four \Lya~lines are detected along this line of sight, two of which
are associated with metal line systems.  There is a strong
high-excitation associated absorption line system at $z=0.5088$ which
includes \hbox{Ly-$\alpha$} and the C~IV and N~V doublets.  The
$z=0.5061$~system consists of \hbox{Ly-$\alpha$} and the strong line of the
C~IV doublet (observed at 2331~\AA). This line is identified as C~IV
even though the weaker line of the doublet is not detected because it
is probable that the strong C~IV~\l1548.20 line at $z=0.5088$~is
blended with C~IV~\l1550.77 from the $z=0.5061$~system.  The incomplete
sample contains the Al~II Galactic ISM line at $1670.70\pm0.19$~\AA~
($W=0.72\pm0.14$~\AA, $SL=4.39$, FWHM$=1.50$~\AA). The line fitting
software did not handle well the strong, non-Gaussian Galactic ISM
Mg~II absorption. The feature at 2799~\AA~is actually the blended
wings of the Mg~II~\ll2796,2803 absorption. There are no candidate
damped \hbox{Ly-$\alpha$} or LLSs along this line of sight (based on the
G190H and G160L observations).

{\bf \boldmath PKS~1136$-$13  ($z_{\rm{em}}$=0.554,\unboldmath ~18 lines)}
\nobreak
There are seven \Lya~lines identified in this spectrum, one of which is
associated with a detected metal line system. For the $z=0.4064$
system, a candidate N~V~\l1238.82 line is detected in the
incomplete sample at $1741.86\pm0.31$~\AA~($W=0.31\pm0.10$~\AA,
$SL=3.10$, FWHM$=1.51$~\AA)

{\bf         3C~263  (RA: 11:37:09.4, DEC:$+$66:04:27, B1950;
\boldmath$z_{\rm{em}}$=0.652)\unboldmath ~ CAT1}       
\nobreak
While our G190H and G270H spectra have been previously published, the
G160L spectrum shown in Figure~2 was not available when CAT1 was
written. We have included the G190H and G270H spectra in Figure~2
along with the new G160L spectrum.

{\bf \boldmath PG~1202$+$281, ($z_{\rm{em}}$=0.165,\unboldmath ~6 lines)}
\nobreak
Only Galactic ISM lines were detected along this line of sight.  The
Fe~II~\l2600.18 line is in the incomplete sample at
$2600.16\pm0.27$~\AA~($W=0.52\pm0.11$~\AA, $SL=4.48$,
FWHM$=2.04$~\AA).

{\bf \boldmath  PG~1206$+$459 ($z_{\rm{em}}$=1.158,\unboldmath ~122 lines)}
\nobreak
Our identifications of the lines in the UV spectra of PG~1206$+$459
are incomplete.  This is in part due to the irregular nature of the
continuum of this quasar and the difficulty of obtaining a good fit.
The Galactic ISM lines, some remarkable extensive metal line systems
(at $z=0.9254$, 0.9277, 0.9342), and a few additional lines have been
successfully identified.  The Lyman$-$limit fitting software also
found a candidate system at $0.928$, a redshift in remarkable
agreement with the narrow line system already mentioned
($z=0.9277$). However, we have not included this candidate LLS in
Table~6 as we consider it not to be a real system, but most likely
just the result of the irregular behavior of the quasar's continuum
(Paper~V).  The majority of the lines in this spectrum have multiple
plausible identifications and we have not yet been able to assign a
``most probable'' identification for all of them.  Most of the lines
in the table~3 marked ``notes'' probably include a contribution from a
line in another system as well. The other lines for which ``notes''
are indicated in table~3 refer to the following: the feature at
1995.68~\AA~is the stronger line of the O~VI doublet at $z=0.9342$ and
the second line of the doublet is in the incomplete sample at
$2006.98\pm0.24$~\AA~with $W=0.213\pm0.06$~\AA, FWHM$=1.50$~\AA, and
$SL=3.74$; the line at 1997.72~\AA~might also include a contribution
from C~II~\l1036.34 in the $z=0.9277$ system; the blended lines at
2160.43 and 2162.00~\AA~might include Fe~III at $z=0.9254$, but are
likely to be produced by multiple lines (including a component of
Fe~II~\l1121.99 at $z=0.9277$); the blended lines at 2386.16 and
2393.38~\AA~are likely to include a contribution from the N~V doublet
at $z=0.9254$, but probably include contributions from other lines as
well; the line at 2985.12~\AA~is a blend of C~IV~\l1550.77 at
$z=0.9254$~and C~IV~\l1548.20 at $z=0.9277$; the line at 3225.37 is
most likely the result of an error in the continuum fit.

{\bf  MC~1215$+$113 (\boldmath$z_{\rm{em}}$=1.396)\unboldmath ~ CAT1}     
\nobreak
We now consider the candidate Lyman$-$limit systems listed in CAT1 not
to be real systems (see Paper~V for discussion) and no longer list
them as part of our Catalogue. See Table~6 for a complete listing of
LLSs found in the survey.

{\bf \boldmath PG~1216$+$069 ($z_{\rm{em}}$=0.334,\unboldmath  ~29 lines)}
\nobreak
The spectrum contains nine identified \hbox{Ly-$\alpha$} lines, including one
of the strongest \hbox{Ly-$\alpha$} absorbers detected in the Key Project
observations. The remarkable line at 1223.34~\AA~($z=0.0063$; 
1890 \hbox{km~s$^{-1}$}) has no detected associated metal lines in the complete
sample; however there are possible detections of Si~II~\l1260.42,
C~II~\l1334.53, and Mg~II~\l2796.35 in the incomplete sample. New
observations with an improved signal-to-noise ratio are needed to test
these candidate identifications and to test whether the broad
\hbox{Ly-$\alpha$} line at 1223~\AA~can be resolved into separate components.
It will be extremely interesting to look at other wavelengths (optical
and radio imaging) for a counterpart to this very low redshift system.

The only acceptable identification for the line at 1741~\AA~is C~IV
associated with the \hbox{Ly-$\alpha$} line at 1367.21~\AA~($z=0.1247$).
\zsearch~formally identifies this line as the stronger component of the
C~IV doublet, but it is possible that the single broad feature found
by the line measurement software is produced by both lines of the
doublet blended together. In addition, the \hbox{Ly-$\alpha$} line at
$z=0.1236$~is close enough in velocity ($<300$ \hbox{km~s$^{-1}$}) that C~IV
from that system might also be blended with the $z=0.1247$ lines.  A
shallow and broad line at 2664~\AA~was removed from the list because
it is produced by an error in the continuum fit.

This line of sight contains a rich spectrum of Galactic ISM lines
including 15 lines in the complete sample and possibly seven
additional lines in the incomplete sample. The tentative Galactic ISM
identifications in the incomplete sample are listed in Table~7.  The
weaker line of the C~IV doublet was not detected in either the
complete or incomplete samples, raising the possibility that the line
at 1548.20~\AA~is actually a \hbox{Ly-$\alpha$} line at $z=0.2736$.

{\bf \boldmath  MARK~205 (RA: 12:19:33.5, DEC:$+$75:35:16.0, B1950;
$z_{\rm{em}}$=0.070)\unboldmath }
\nobreak
Although too low a redshift to have been included in the original Key
Project sample of targets, MARK~205 was one of the GTO observations of
Bahcall and is now included in this summary for the sake of
completeness.  The spectrum and line identifications for this object
were presented by Bahcall \etal~(1992b).

{\bf           3C~273 (RA: 12:26:33.3, DEC:$+$02:19:43, B1950;
\boldmath$z_{\rm{em}}$=0.158)\unboldmath ~ CAT1}       
\nobreak
Our FOS spectra of this quasar were first analyzed by Bahcall
\etal~(1991a,b) and reanalyzed with the Key Project analysis software
in CAT1.

{\bf \boldmath PG~1241$+$176 ($z_{\rm{em}}$=1.273,\unboldmath ~41 lines)}
\nobreak
This spectrum contains 19 \Lya~lines (two are associated with detected
metal lines), and there is a remarkably strong candidate O~VI doublet
at a redshift close to that of the quasar. The only additional lines
in the system are \Lya~and \Lyb. Other features that might have been
expected (\eg~the N~V and C~III) are not present.  However, it is
possible that there is N~V~absorption present in the spectrum and
that it has been removed by our choice of continuum fit (see
Figure~2). If the lines comprising the candidate O~VI doublet at
$z=1.2720$~are misidentified, the alternative identifications for the
lines are \Lya~at $z=0.9286$~and $0.9395$. The line at
2344.53~\AA~also includes a small contribution from the Galactic ISM
Fe~II~\l2344.21~line.  The line at 3205.24~\AA~is probably the result of
an imperfect continuum fit.  The feature at 2532~\AA~is the result of
a flat-fielding residual and an error in the continuum fit. The
\Lya~line at 2244~\AA~might be the result of the continuum fitting.

{\bf B2~1244$+$32B (\boldmath$z_{\rm{em}}$=0.949)\unboldmath ~ CAT1}

{\bf \boldmath PG~1248$+$401 ($z_{\rm{em}}$=1.030,\unboldmath ~88 lines)}
\nobreak
This line of sight includes four metal line systems (two very
extensive) in addition to the 37 \Lya~(four of which are associated
with the metal line systems) and eight Galactic ISM lines.  The metal
line system at $z=0.8553$ includes a broad \Lya~feature that was split
by the fitting software into two components. Absorption by \Lya, \Lyb,
\Lyg, \Lyd, \Lye, C~II, C~III, C~IV, N~III, N~V, O~VI, Fe~III, and
Si~III is observed in this system.  The second extensive system, at
$z=0.7732$, includes lines from \Lya, \Lyb, \Lyg, \Lyd, \Lye, C~II,
C~III, C~IV, N~II, N~III, N~V, O~VI, Fe~III, Si~II, Si~III, and Si~IV.
The Fe~III line in this system was found as part of a candidate system
at $z=0.7734$ that included many, but not all of the lines in the
candidate system at $z=0.7732$ -- while formally the wavelength
agreement is insufficient for this line to be identified with this
system, it is likely to be the correct identification and that the
redshift is slightly in error.  The line at 1922.34~\AA~is listed in
the table as C~II in the $z=0.8553$ system, but is blended with the
N~II line from the $z=0.7732$ system.  The line at 1924.96~\AA~is
listed as O~VI in the $z=0.8553$ system, but there is a small
probability that this line is blended with N~II in the $z=0.7660$
system.  The candidate metal line system at $z=0.7760$ is of
particular interest because there is no detected C~IV doublet to
accompany the detected O~VI lines.  This is an excellent candidate to
be an absorber produced by collisionally excited gas, perhaps
associated with a group or cluster of galaxies.  If the candidate O~VI
lines in the $z=0.7760$ system are not O~VI (and identified lines in
this system only include \Lya, \Lyb, and the O~VI doublet), then they
are most probably \Lya~lines at $z=0.507$ and 0.516.

Alternative identifications for the 1887.55 and 1890.21~\AA~lines are
the lines in a C~IV doublet at $z=0.2190$. Galactic ISM
Fe~II~\l2260.78 and Fe~II~\l2382.77~are probably blended with the
\Lya~lines at respectively 2259.30 and the blend at 2383, 2387~\AA.

{\bf \boldmath PKS~1252$+$11 ($z_{\rm{em}}$=0.870,\unboldmath ~49 lines)}
\nobreak
There are 24 identified \Lya~lines in this spectrum, including one
that may be associated with a detected metal line system.  The
O~VI~\l1037.62 line in the $z=0.6395$ system is in the incomplete sample
at 1702.45$\pm$0.22~\AA~with $W=0.45\pm0.10$~\AA, FWHM$=1.51$~\AA, and
a $SL=1.51$.  Any Mg~II associated with this system should be
observable from the ground.  The line at 2027.19~\AA~is a very broad
feature produced by a real feature and an uncertain continuum fit. The
features at 2063.78~\AA, 2214.15~\AA, and 2637.55 \AA~similarly appear
to be blends of weaker features and/or slight errors in the continuum
fit. The 1780.14 and 1878.12~\AA~lines, identified as \Lyb~at the
$z=0.7357$ and \Lya~at $z=0.5449$, possibly include contributions from
respectively \Lyg~and \Lyb~at $z=0.8307$.  Alternative
identifications for the \Lya~lines at 2266.75 and 2276.58~\AA~are as
the Al~III doublet at $z=0.2222$.

{\bf PG~1254$+$047  (\boldmath$z_{\rm{em}}$=1.024)\unboldmath ~ }
\nobreak
This is one of the six BAL quasars observed by the Key Project and
Bahcall GTO program.  The spectrum is displayed in Figure~2, but
the analysis of this spectrum will be presented in Turnshek \etal~1998.

{\bf \boldmath B201~1257$+$57 ($z_{\rm{em}}$=1.375,\unboldmath ~25  lines)}
\nobreak
This spectrum contains 13 \Lya~lines, three of which are associated
with metal lines.  The \Lyd~line of the associated absorption system
(at $z=1.3799$) is blended with the line identified as Galactic ISM
Fe~II~\l2260.78.  Candidate Fe~II lines from the Galactic ISM are in
the incomplete sample at 2248.60$\pm$0.29~\AA, $W=0.70\pm0.16$~\AA,
FWHM$=2.04$~\AA, $SL=4.19$ and 2375.02$\pm$0.35~\AA,
$W=0.45\pm0.12$~\AA, FWHM$=2.04$~\AA, $SL=3.48$.  The Galactic ISM
Fe~II~\l2382.77 line is anomalously strong, relative to the other
Galactic ISM lines, and probably includes a blend from another line.

{\bf  PG~1259+593   (\boldmath$z_{\rm{em}}$=0.472)\unboldmath ~  CAT1}

{\bf \boldmath PKS~1302$-$102 ($z_{\rm{em}}$=0.286,\unboldmath ~39 lines)}
\nobreak
A total of 21 Galactic ISM lines was detected in the complete sample
including absorption from N~I, Si~II, S~II, O~I, C~II, C~IV, Si~IV, Al~II,
Mg~I, Fe~II, Mn~II, and possibly Zn~II.  In the incomplete sample
S~II~\l1253.81~at $1253.70\pm0.19$~\AA~is observed with an equivalent width
of $0.22\pm0.06$ \AA~and $SL=3.49$, and C~IV~\l1550.77 at $1551.01\pm0.16$
\AA~with an equivalent width of $0.11\pm0.03$~\AA~and $SL=4.05$.
Both of the Galactic ISM C~IV lines occur on top of a strong quasar
emission line that makes their measured equivalent widths uncertain
due to the uncertainty in the continuum fit. Fe~II~\l2249.88 is seen as a
broad blend located at 2249.3 \AA, and Mg~I (Zn~II) is in the
incomplete sample (2026.89$\pm$0.28 \AA~with $W=0.15\pm0.05$~\AA,
FWHM$=1.51$~\AA, and $SL=3.30$).

There are 14 \Lya~lines identified in the spectrum of this object.
The line at 1283.96~\AA, fitted by the software as a single
``resolved'' line, is probably a blend of more than one line. If we
force a fit with two lines, the individual components have centers at
approximately 1283.8 and 1284.8 \AA.  The single line in the complete
sample may be composed of the \hbox{Ly-$\beta$} line from a system at
redshift 0.2521 and a \hbox{Ly-$\alpha$} line at a redshift of 0.0562; higher
resolution spectroscopy is needed to resolve this question.  The
\hbox{Ly-$\alpha$} line at 1489.49~\AA~is strong and broad.  Any associated
C~IV was too weak to appear in either the complete or incomplete
samples.

{\bf TON 153 (RA: 13:17:34.3, DEC:$+$27:43:52, B1950;
\boldmath$z_{\rm{em}}$=1.022)\unboldmath ~ CAT2} 
\nobreak
Note the slight revision (see Table~6) of the properties of the
Lyman$-$limit system first reported in CAT1. These are as a result of
the changes made to the software measuring the LLSs and not due to any
change in the reduction of the spectrum.

{\bf \boldmath PG~1333$+$176 ($z_{\rm{em}}$=0.554,\unboldmath ~14 lines)}
\nobreak
Three \Lya~lines appear in this spectrum, one of which is associated
with a C~IV system at a redshift 0.3458. The C~IV~\l1550.77 line is
present in the incomplete sample with an equivalent width of
$0.21\pm0.05$~\AA, an observed wavelength of $2087.31\pm0.25$~\AA, and
a $SL=3.69$.  \hbox{Ly-$\alpha$} for this system is in the complete sample
and Si~IV at 1893~\AA~is in the incomplete sample. The Si~IV
absorption observed at 1887~and 1893~\AA~occurs on top of one of the
quasar's emission lines and the continuum placement affected the
detection of the Si~IV \l1393 at 1887~\AA, which is clearly present on
inspection of the spectrum.

At the redshift of the \hbox{Ly-$\alpha$} line at 1861.9~\AA~($z=0.5316$)
there are no detected metal lines in the complete sample; however,
there is a line in the incomplete sample that matches the expected
wavelength of C~IV~\ll1548.20,1550.77 and would be blended with Galactic
Fe~II~\l2374. Formally, we identify the 1861.9~\AA~line as a \hbox{Ly-$\alpha$}
absorber with no associated metals, but it is worth remembering that
such statements have the implied qualifier: ``no metal lines strong
enough to be included in the complete sample''.

Galactic ISM lines in the complete sample are detected from Fe~II,
Mg~II, and Mg~I. Al~II~\l1670.79 is present in the incomplete sample
with an observed wavelength of 1670.78~\AA, an equivalent width of
$0.46\pm0.11$~\AA, and a significance level of 4.00.

There is some evidence that both the 2919 and 2958~\AA~lines could
arise from imperfect flat-field correction, but is not sufficient to
allow us to identify these features as flat-fielding artifacts.

{\bf \boldmath PG~1338$+$416 ($z_{\rm{em}}$=1.219,\unboldmath ~39 lines)}
\nobreak
This spectrum has a lower signal-to-noise ratio than the vast majority
of the Key Project spectra. Combined with the presence of several
strong blends or broad absorption features, the poor signal made
fitting the continuum of this spectrum particularly difficult and the
resulting line measurements particularly uncertain.  Despite these
problems, this line of sight is unusual for the number of strong \Lya~
lines observed in the spectrum.  In total 22 \Lya~lines are
identified, two of which have associated metal lines. An extensive
metal line system is identified at $z=1.2152$.  Only one Fe~II line
(1144~\AA) is observed in this system, and the line (observed at
2535.85~\AA) might alternatively be part of a C~IV doublet at
$z=0.6351$.  The line at 2400~\AA~might include a contribution from
N~II~in the $z=1.2152$ system.  The second line of the C~IV doublet
at $z=1.0778$ is in the incomplete sample at 3222.17$\pm0.37$~\AA~with
$W=0.34\pm0.10$~\AA, FWHM$=2.04$~\AA, and $SL=3.30$.  The strong line
at 2383.17~\AA~must include some Galactic ISM Fe~II in addition to the
identified \Lya~line.  The most probable alternative identifications
for the lines comprising the candidate C~IV doublets at $z=0.6213$ and
$z=0.6863$ are as \Lya~lines.  Galactic ISM Fe~II~\l2586.65, which is
stronger than expected, probably includes a contribution from another
line (\eg~\Lya~line at $z=1.1274$).

{\bf B2~1340$+$29  (\boldmath$z_{\rm{em}}$=0.905)\unboldmath ~ CAT1 }

{\bf 4C~53.28 (RA: 13:47:42.7, DEC:$+$53:56:09, B1950;
\boldmath$z_{\rm{em}}$=0.976)\unboldmath ~ CAT1 }

{\bf PG~1352$+$011 (\boldmath$z_{\rm{em}}$=1.121)\unboldmath ~ CAT2 }

{\bf \boldmath PKS~1354$+$19 ($z_{\rm{em}}$=0.720,\unboldmath  ~23 lines)}
\nobreak
Eight \Lya~lines are identified in this object's spectrum, one
of which is associated with an extensive metal line system at
$z=0.4563$ that almost certainly corresponds to the LLS detected in
our G160L spectrum (see CAT1 and Paper~V) at $z=0.45$; the properties
of the LLS are listed in Table~6. Absorption by Si~II~\l1190.42
associated with this system is detected in the incomplete sample at
$1733.28\pm0.21$ ($W=0.47\pm0.10$~\AA, $SL=4.34$, FWHM$=1.51$~\AA).

Galactic ISM Fe~II~\l2382.77 is in the incomplete sample at
$2382.89\pm0.33$~\AA~($W=0.43\pm0.11$~\AA, $SL=3.73$,
FWHM$=2.04$~\AA).  Contrary to expectations, the equivalent width of
Galactic ISM Mg~II~\l2803 is observed to be stronger than the
Mg~II~\l2796 line by 0.26~\AA.  If the 2803~\AA~line is contaminated
by high velocity Mg~II~\l2796 absorption, the implied velocity is
approximately $+800$~\hbox{km~s$^{-1}$}.  Such extreme high velocity clouds
are unlikely.

{\bf \boldmath PG~1407$+$265 ($z_{\rm{em}}$=0.944,\unboldmath ~82 lines)}
\nobreak
Our \HST~observations of this quasar have been previously published by
McDowell \etal~(1995) when they discussed the unusual emission-line
properties of this radio quiet quasar. As they reported, the vast
majority of the emission lines have comparatively small equivalent
widths and cover a very wide range in redshift (over 10,000
\hbox{km~s$^{-1}$}).  Because different emission lines yielded quite
different redshifts, McDowell et al.~quote a redshift for the quasar
of $z=0.94\pm0.02$.  Given our detection of a metal line absorption
system at $z=0.9566$, the systemic redshift of the quasar is likely to
be at least as large as 0.95.

Alternative identifications for the lines at 1855.06 and
1858.13~\AA~are as a C~IV doublet at $z=0.1984$. No \Lya~absorption is
detected in association with the candidate C~IV doublet at
$z=0.4115$. Alternative identifications of this doublet are as
\Lya~lines at $z=0.7976$~and 0.8006. If the candidate N~V doublet
with associated \Lya~at $z=0.4053$ identifications are correct, then
the strength of any associated C~IV absorption is less than
0.12~\AA~(for the stronger line of the doublet).  The 2344.30~\AA~line
is expected to include a contribution from Galactic ISM Fe~II, but has
been identified as
\Lya~because of its large equivalent width.  The line at 2030.67~\AA~
has numerous possible identifications, and we have chosen the one that
would place the line in the most extensive of the candidate systems;
however, the identification of this line should be considered quite
uncertain.

{\bf PG~1411$+$442  (\boldmath$z_{\rm{em}}$=0.089)\unboldmath ~ }
\nobreak
PG~1411$+$442 has too low a redshift to have been included in the
original Key Project sample of targets. It was one of the GTO targets
of Bahcall and is included in this paper for completeness
since our catalogue contains the results of the analysis of all of the
objects observed as part of both the Key Project and the Bahcall GTO
program.  The spectrum is presented in Figure~2.  This is one of the
six BAL quasars observed by the Key Project and Bahcall GTO program.
The analysis of this spectrum will be presented in Turnshek \etal~1998.

{\bf \boldmath PG~1415$+$451 ($z_{\rm{em}}$=0.114,\unboldmath ~6  lines)}
\nobreak
In the spectrum of PG~1415$+$451 we detect in the complete sample only
the familiar Galactic ISM lines of Fe~II, Mg~I, and Mg~II.  There are
three Galactic ISM lines in the incomplete sample from Fe~II (at
$2250.77\pm0.45$ \AA, $W=0.33\pm0.12$ \AA, $SL=3.50$, FWHM$=2.40$
\AA~and $2374.70\pm0.28$ \AA, $W=0.36\pm0.08$ \AA, $SL=4.36$,
FWHM$=2.04$~\AA) and Mg~I (at $2852.83\pm0.45$~\AA,
$W=0.25\pm0.09$~\AA, $SL=3.77$, FWHM$=2.53$~\AA).

{\bf MC~1415$+$172 (\boldmath$z_{\rm{em}}$=0.821)\unboldmath ~ CAT1 }

{\bf \boldmath PKS~1424$-$11 ($z_{\rm{em}}$=0.805,\unboldmath ~28 lines)}
\nobreak
There are 15 \Lya~lines identified in the spectrum of this object, one
of which is associated with a detected C~IV doublet at $z=0.6553$.
This metal line system's redshift is in agreement with the measured
LLS redshift, $z=0.65$, evident in the G160L spectrum (see Table~6 for
measured properties of the system).  In some of our other spectra
there is a flat field feature at 2120~\AA, but not strong enough to
produce the feature observed in this spectrum.  There is a candidate
C~IV doublet at $z=0.0965$ whose identification lacks any additional
supporting evidence (for example the presence of other lines in the
system), and these lines (1697.87 and 1700.23~\AA) have been
identified as \Lyb~lines in other systems.  If these lines are not
caused by \Lyb~or C~IV, they most likely are \Lya~lines. Similarly,
there is a candidate C~IV system at $z=0.3485$, consisting of lines at
2087 and 2091~\AA~currently identified as being produced by
\Lya~absorption; however there are no additional lines detected that
support the existence of the candidate C~IV system.

{\bf \boldmath S4~1435$+$63 ($z_{\rm{em}}$=2.060,\unboldmath ~73 lines)}
\nobreak
The heavy blending of the lines in this spectrum and limited observed
wavelength coverage (exacerbated by the presence of a LLS)
made the fitting of the continuum highly uncertain and the
possible identifications for some lines too numerous to allow complete
identification of the line list at this time.  We do include
identifications of the strong hydrogen lines and tentative
identifications of a few metal lines that might be associated with the
LSS observed at $z=1.9254$ (see Table~6 for additional properties of
the LLS, measured by the Lyman$-$limit search software to have
$z=1.92$), as well as some of the other strong lines in the spectrum.
Three of the Lyman series lines associated with the LLS are heavily
blended with other lines (\Lyd, \Lye, and \Lyl).

There might be a slight error in the zero point of the wavelength
scale of this spectrum, since the Galactic ISM Fe~II lines that we
normally use to set the zero point were not observable due to the
LLS. Strong lines that are likely to have been produced by Mg~II and
Mg~I in the Galactic ISM are present at an observed redshift
of~0.0003. If these identifications are correct and the lines are due
to gas in the Galaxy, the true redshift of these lines is likely to be
closer to zero and all of the wavelengths presented in the table of
identifications need to be shifted by approximately $-0.9$ to
$-1.0$~\AA. However, the redshift of the LLS measured in our G160L
spectrum is in excellent agreement with the redshift measured in the
system's narrow lines, suggesting that in fact the lines at 2797.28
and 2804.78~\AA~might have an alternative identification or be blended
with other lines that are displacing the line centers. Higher spectral
resolution observations would remove this uncertainty.

{\bf \boldmath PG~1444$+$407 ($z_{\rm{em}}$=0.267,\unboldmath ~9 lines)}
\nobreak
All the identified absorption lines are from the Galactic ISM.
Planned G130H observations of this object had to be dropped when the
original time allocation was reduced. Galactic ISM Mg~I~\l2852 is in the
incomplete sample at $2852.90\pm0.33$~\AA~($W=0.21\pm0.05$~\AA,
$SL=3.74$, FWHM$=2.04$~\AA).

{\bf        B2~1512$+$37 (\boldmath$z_{\rm{em}}$=0.370)\unboldmath ~ CAT1}

{\bf \boldmath PG~1538$+$477 ($z_{\rm{em}}$=0.770,\unboldmath ~81 lines)}
\nobreak
This line of sight contains 31 identified \Lya~lines, including four
lines associated with detected metal lines.  Two extensive metal line
systems (at $z=0.7300$, 21 lines, and $z=0.7705$, 11 lines), are also
present along this line of sight.  The $z=0.7300$ system is associated
with a LLS detected in the G160L spectrum (see Table~6 for the
properties of the LLS as determined by the Lyman$-$limit search
software).  The weaker line of the N~V doublet in the $z=0.7300$~
system is heavily blended with the \Lya~line at $z=0.7705$, and the
measured properties of the line are very uncertain.  The $z=0.4863$
system consists of a \Lya~line at 1806~\AA~and a candidate C~IV
doublet with its two lines at 2301.14~\AA~and blended with the line at
2305~\AA, identified as O~I at $z=0.7705$.  The \Lya~line at
1622.40~\AA~occurs in a region of the spectrum with poor flat-fielding
and flux calibration; we suspect that the true uncertainties in the
measured line parameters are larger than the tabulated values.  The
weaker line of the candidate C~IV doublet at $z=0.5166$ is blended
with a strong flat-fielding residual at 2353~\AA.  The expected \Lyb~
line from the $z=0.7071$ system would be blended with the line at
1750.90~\AA, which is identified as \Lya.  The \Lya~line at
1866.98~\AA~is possibly affected by a flat-fielding residual and is
heavily blended with the \Lya~line at 1868.51~\AA.  The lines at
1890.06 and 1892.90~\AA~(1928.52 and 1931.07~\AA), identified as
\Lya~lines could be, under our identification rules, listed as a C~IV
doublet at $z=0.2207$~($z=0.2456$). This spectrum provides some very
weak lines ($W<0.20$~\AA) for the complete sample line list. The
reality of these weak features is uncertain since our ability to
identify flat-field residuals in this spectrum is limited to lines of
this strength.  Note that redshift of this quasar has been revised
since the value we used for the analysis in this paper (from the
V\'eron-Cetty \& V\'eron 1991 catalogue) and is now believed to be
0.772 (see reference in V\'eron-Cetty \& V\'eron 1996).

{\bf \boldmath 3C~334.0 (RA: 16:18:07.3, DEC: $+$17:43:29.0, B1950;
$z_{\rm{em}}$=0.555,\unboldmath ~27 lines)}
\nobreak
Eleven \hbox{Ly-$\alpha$} absorption lines are identified in the complete
sample. There is no evidence in the G160L spectrum for any damped
\hbox{Ly-$\alpha$} or LLSs in the region between 1200 and 1600~\AA.

This spectrum possesses a rich collection of Galactic ISM lines
including 14 lines in the complete sample from the following: Al~II,
Al~III, Zn~II, Mg~I, Mg~II, Fe~II, and Mn~II.  The large FWHM of the
Al~III~\l1854.72 and Mn~II lines suggests that they are blended
with other lines. Weaker lines of Al~III and Mn~II are found in the
incomplete sample (Al~III~\l1862.79~at 1862.73~\AA~with an equivalent width
of $0.17\pm0.04$~\AA, $SL=3.84$; Mn~II~\ll2576.88,2594.50 with line centers,
equivalent widths, and significance levels of 2577.9 and 2594.57~\AA~;
$0.25\pm0.069$~\AA~and $0.29\pm0.069$~\AA; and 3.66 and 4.04).

{\bf PG~1634$+$706 (\boldmath$z_{\rm{em}}$=1.334)\unboldmath ~ CAT2 }
\nobreak
After the line list derived from our G270H spectrum of this object
was published in CAT2, we obtained the G190H spectrum of Impey
\etal~(1995, 1996) from the HST archive and processed the spectrum in a manner
consistent with our reductions of the other spectra. We measured the
properties of the LLSs identifiable in the G190H spectrum and added
their properties to the systems listed in Table~6.  Both of the LLSs
can be associated with extensive metal line systems first reported in CAT2.
Since the Al~II~\l1670.79 Galactic ISM line was not detected in the
G190H spectrum (due to the LLSs) we could not check the zero-point of
the wavelength scale for that portion of the spectrum.

{\bf PKS~1656$+$053 (\boldmath$z_{\rm{em}}$=0.879)\unboldmath ~ CAT1 }

{\bf PG~1700$+$518 (\boldmath$z_{\rm{em}}$=0.290)\unboldmath ~ }
\nobreak
This is one of the six BAL quasars observed by the Key Project or
Bahcall GTO program.  The spectrum is displayed in Figure~2, but
the analysis of this spectrum will be presented in Turnshek \etal~1998.

{\bf        3C~351 (RA: 17:04:03.5, DEC:$+$60:48:31, B1950;
\boldmath$z_{\rm{em}}$=0.371)\unboldmath ~     CAT1}

{\bf \boldmath PG~1715$+$535 ($z_{\rm{em}}$=1.929,\unboldmath ~32 lines)}
\nobreak
The very low signal-to-noise ratio of the spectrum, heavy blending of
the lines, and limited observed wavelength coverage made the fitting
of the continuum uncertain and the possible identifications for some
lines too numerous to allow complete identification of the line list.
Some lines associated with the LLS at $z=1.6333$ were identified (note
that the Lyman$-$limit search software found a slightly different
redshift based on measurement of the general depression of the
continuum near 2400~\AA, $z=1.64$; the method used in Paper~V finds
$z=1.626$). The properties of the LLS as measured by the Lyman$-$limit
search software are listed in Table~6. The stronger of the N~V doublet
lines for the $z=1.6333$ system is in the incomplete sample at
3262.15$\pm0.34$~\AA~with $W=0.84\pm0.25$~\AA, FWHM$=2.42$~\AA, and
$SL=4.40$.  Galactic ISM Fe~II~\l2382.77 is in the incomplete sample at
2382.27$\pm0.34$~\AA~with $W=1.32\pm0.43$, FWHM$=2.30$~\AA, and
$SL=3.92$.

{\bf \boldmath PG~1718$+$481 ($z_{\rm{em}}$=1.084,\unboldmath ~88 lines)}
\nobreak
The line identifications for this spectrum are not complete.  There
are several metal line systems, including two C~IV doublets at
$z=1.0323$ and 1.0872.  The 1037~\AA~line of the O~VI doublet in the
$z=0.8929$ system is in the incomplete sample at
1964.12$\pm0.21$~\AA~with $W=0.19\pm0.04$~\AA, FWHM$=1.51$~\AA, and
$SL=4.28$. The line at 1801.34~\AA~might be N~III at $z=0.8200$.  The
broad feature at 1866.63~\AA~might include \Lyb~at $z=0.8200$ and
Fe~II at $z=0.7012$. The absorption at 2517.30~\AA~might include a
contribution from Si~III at $z=1.0872$.  The Galactic ISM
Mg~II~\l2796.35 line has a negative velocity wing suggesting high
velocity gas.

{\bf    H~1821$+$643  (\boldmath$z_{\rm{em}}$=0.297)\unboldmath ~   CAT1}

{\bf \boldmath 4C~73.18 (RA: 19:28:49.4, DEC: $+$73:51:45, B1950;
$z_{\rm{em}}$=0.302,\unboldmath  ~12 lines)}
\nobreak
There are eleven Galactic ISM lines in the complete sample.
Originally far UV observations using the G130H grating were planned
for this object, but reductions in subsequent time allocations
necessitated dropping the G130H observations from the program. The
broad line at 1884~\AA~might be a C~IV doublet that was not well
resolved because of the low signal-to-noise ratio in this spectrum.

\vbox{
{\bf PG~2112$+$059 (\boldmath$z_{\rm{em}}$=0.457)\unboldmath ~ }
\nobreak
This is one of the six BAL quasars observed by the Key Project and
Bahcall GTO program.  While the continuum of this particular BAL
is not as difficult to fit, we defer the analysis of this spectrum to
our discussion of the other BAL quasars, which will be presented in
Turnshek \etal~1998. The spectrum is displayed in Figure~2.
}

{\bf \boldmath PKS~2128$-$12 ($z_{\rm{em}}$=0.501,\unboldmath ~28 lines)}
\nobreak
The spectrum contains a rich low-excitation metal line system at
$z=0.4296$ that includes twelve lines in the complete sample (C~II,
C~IV, O~I, N~I, Si~II, Si~III, Si~IV, Fe~II, Al~II, and
\hbox{Ly-$\alpha$}). The \hbox{Ly-$\alpha$} of this system is strong (rest
equivalent width of 2.9~\AA). There are five additional \Lya~lines in
this spectrum, none that are associated with identified metal
lines.

The weak lines on top of the strong quasar emission line between
1800~\AA~and 1824~\AA~have larger systematic uncertainties than the
quoted formal uncertainties as it is difficult to fit the continuum in
this region.

{\bf     PKS~2145$+$06  (\boldmath$z_{\rm{em}}$=0.990)\unboldmath ~ CAT1 }

{\bf \boldmath PKS~2243$-$123, ($z_{\rm{em}}$=0.630,\unboldmath  ~26 lines)}
\nobreak
Ten \Lya~lines are identified in the complete sample.  The line at
1822~\AA~is possibly an artifact of the continuum fit.  Galactic ISM
Al~II~\l1670.79 is in the incomplete sample at 1670.76$\pm0.21$~\AA,
with $W=0.33\pm0.07$~\AA, FHWM$=1.51$~\AA, and a $SL=4.44$.  The
non-Gaussian wing to the Mg~II~\l2796.35 absorption produces the
component near 2793~\AA.

{\bf \boldmath PKS~2251$+$11 ($z_{\rm{em}}$=0.323,\unboldmath ~38 lines)}
\nobreak
In this paper we combine a new G130H spectrum of this object with the
previously presented G190H, G270H, and G160L spectra (CAT1). All of
these data have been analyzed with the current version of the line
fitting and identification software and the measurements and
identifications presented in this paper replace those presented in
CAT1.  The G130H data confirm the metal line system identified in CAT1
at $z=0.3256$ from only the four lines in the G190H and G270H spectra
(\hbox{Ly-$\alpha$}, C~IV doublet, and N~V~\l1238.82).  The new lines in this
system found in the G130H spectrum are \hbox{Ly-$\beta$} and the O~VI
doublet.  The G130H data also resolve the possible damped \hbox{Ly-$\alpha$}
system noted in CAT1 (based on our lower resolution G160L data) into a
combination of weaker absorption features and quasar emission.  The
continuum between 1300 to 1400~\AA~was particularly difficult to
fit.

Differences between our current analysis of the G190H and G270H
spectra and that appearing in CAT1 include the following: the line
previously identified as Galactic ISM Fe~II~1608.08 is now \Lya~at
$z=0.3236$; one unidentified line from the CAT1 list dropped out of
the complete sample; four new lines appear in the revised complete
sample, including three unidentified lines and Si~IV~\l1393.76 in the
$z=0.3256$ system.  The lines at 1841 and 1853~\AA~were selected by
\zsearch~as a candidate Si~IV absorption doublet at $z=0.3214$, but no
other lines are found at this redshift to support this identification.
The 1853~\AA~line might be partially produced by Galactic ISM
Al~III~\l1854.72.  The identification of 1393~\AA~as Galactic ISM Si~IV
is weakened by the lack of any Galactic ISM lines from C~IV or N~V in
the complete sample and the absence of the 1402~\AA~line in the
incomplete sample.  The line at 1453.30~\AA~appears to be a blend of
several lines that the software fit as a single line.  The \hbox{Ly-$\alpha$}
identification at 1233.80~\AA~should be considered quite uncertain,
despite the high measured significance level of the lines, because the
low signal in this region makes the systematic error in the continuum
fit large.  The total number of identified \Lya~lines in this spectrum
is 15, with one of these lines associated with the previously
discussed metal line system at $z=0.3256$.

{\bf 3C 454.3 (RA: 22:51:29.6, DEC:$+$15:52:54, B1950;
\boldmath$z_{\rm{em}}$=0.859)\unboldmath ~ CAT1}

{\bf \boldmath PKS~2300$-$683 ($z_{\rm{em}}$=0.512,\unboldmath ~7 lines)}
\nobreak
Only seven lines are in the complete sample from the
G190H and G270H spectra of this object.  There are two \Lya~lines
detected, one of which lies about 2000 \hbox{km~s$^{-1}$} shortward of the
quasar emission-line redshift.  The continuum was adjusted
subjectively to fit the quasar emission lines in the wavelength
regions 2340 to 2390~\AA~and 1800 to 1890~\AA. As a result, absorption
lines in these regions probably have larger errors than the tabulated
values and some lines might have been added or removed from the
complete sample with a slightly different continuum fit.  Galactic ISM
lines of Fe~II and Mg~II are identified.  Additional Galactic ISM
Fe~II~\ll2260,2586 lines can be found in the incomplete sample
with line centers of 2260.78 and 2585.97~\AA, equivalent widths of
$0.22\pm0.06$~\AA~and $0.45\pm0.11$~\AA, and significance levels of
3.65 and 3.99.  The G160L spectrum allowed a search for Lyman$-$limit
or damped systems between 1200 and 1600~\AA; no significant features
were detected although there is weak evidence of a LLS near
1310~\AA.

{\bf \boldmath PG~2302$+$029 ($z_{\rm{em}}$=1.052, \unboldmath ~80 lines)}
\nobreak
The HST spectra of this quasar were previously published and some of
its absorption line systems discussed in Jannuzi~\etal~1996.  The
quasar emission redshift we list is from Steidel \& Sargent 1991.  The
spectrum includes a high-ionization broad absorption line system at a
redshift of $z=0.695$~and a narrow line system at 0.7160 which may be
caused by either ejected or intervening material.  Additional
discussion of this object's unusual absorption systems is presented by
Hamann (1997).  In this paper we present measurements for the complete
sample of lines in the spectrum.  In total 35 \Lya~lines are
identified, five of which are associated with identified metal line
systems. The lines at 1932.49 and 1934.56~\AA~are heavily blended,
resulting in a large uncertainty in the measurement of the equivalent
widths of each line. A single fit to the feature near 1933~\AA~would
be justifiable. This would produce better agreement between the
expected wavelength of the \Lya~in the $z=0.5904$~system and the
measured line.  If the line at 2062.62~\AA~is not due to Galactic ISM
Zn~II, then it is \Lya~at $z=0.6967$.  The identification of the line
at 2652.77~\AA~is uncertain, but it is plausibly caused by additional
C~IV absorption in the $z=0.695$ system.  The Galactic ISM
Mg~II~\l2796.35 line has a positive velocity wing implying the
presence of high velocity gas.

{\bf \boldmath PKS~2340$-$036 ($z_{\rm{em}}$=0.896,\unboldmath ~79 lines)}
\nobreak
There are 31 \Lya~lines identified in this spectrum, of which seven
are associated with metal line systems.  Six candidate C~IV doublets
are detected along this line of sight at the following redshifts:
0.1509, 0.1691, 0.4088, 0.4212, 0.4621, and 0.6841. Note that there is
a significant probability that some of these are false systems
produced by the chance combination of other lines (see \S3.4).  If the
lines at 1781 and 1784~\AA~are not a C~IV doublet at $z=0.1509$, the
1781~\AA~line is probably C~III at $z=0.8238$ and the line at
1784~\AA~line would be \hbox{Ly-$\beta$} corresponding to the \hbox{Ly-$\alpha$}
line at $z=0.7401$.  If the lines at 1809 and 1813~\AA~are not a C~IV
doublet at $z=0.1691$, the 1809~\AA~line is probably \hbox{Ly-$\alpha$} at
$z=0.4888$ and the 1813~\AA~line would be \hbox{Ly-$\beta$} at $z=0.7679$.
Absorption by the weaker line of the Si~IV doublet in the $z=0.4212$
system is detected in the incomplete sample at
$1993.56\pm0.22$~\AA~($W=0.24\pm0.06$~\AA, $SL=4.23$,
FWHM$=1.51$~\AA).

Systems with candidate O~VI absorption are observed at redshifts of
0.6841 and 0.8238.  The O~VI doublet identifications for the
$z=0.8238$ system are uncertain because several other systems also
have lines expected at the wavelength of these lines (\eg~1893~\AA~
might be partially \hbox{Ly-$\beta$} at $z=0.8463$). The O~VI identification
is strengthened by the possible identification of C~IV~\l1550.77 for
this system in the incomplete sample ($2823.18\pm0.38$~\AA,
$W=0.25\pm0.08$~\AA, $SL=9.35$, FWHM$=2.04$~\AA). Even if the O~VI
identifications are correct, it is likely that the \hbox{Ly-$\beta$} line at
$z=0.8463$ contributes to the broad absorption observed at 1893~\AA.
Systems like the one at $z=0.8238$ are of particular interest because
they might be tracing material that has been collisionally excited by 
gas associated with a group or cluster of galaxies.

The line at 1773~\AA~identified as \hbox{Ly-$\gamma$} at $z=0.8238$ probably
includes a contribution from \hbox{Ly-$\beta$} at $z=0.7296$.  The lines at
2044 and 2047~\AA~(both identified as \hbox{Ly-$\alpha$} at $z=0.6817$ and
0.6841) are blended together.  There are three features that appear
to be real lines in the reddest portion of the G270H spectrum that 
remain unidentified (3125, 3131, and 3138~\AA).

The Galactic ISM Mg~II~\ll2796.35,2803.53 lines have similar profiles
and both appear to have high velocity wings on their blue sides which
were fitted as separate components by the software.  The incomplete
sample includes absorption by Galactic ISM Fe~II at
$2260.71\pm0.26$~\AA~($W=0.11\pm0.03$~\AA, $SL=3.53$,
FWHM$=1.50$~\AA).

{\bf \boldmath PKS2344$+$09 ($z_{\rm{em}}$=0.677,\unboldmath ~30 lines)}
\nobreak
This spectrum contains 18 \Lya~lines, one of which is associated with
the metal line system at $z=0.4368$. \zsearch~identified a candidate
system at $z=0.3568$ including \Lya, Si~II~\l1260.42, and
O~I~\l1302.17, but we have identified all of these lines as \Lya~lines
pending the identification of additional supporting lines.  For
example, if associated Mg~II lines were to be observed at
approximately 3794~\AA, the identifications listed above would be more
likely.  The Galactic ISM Mg~II lines are very broad and likely
contain unresolved high velocity components.

{\bf \boldmath PKS~2352$-$342 ($z_{\rm{em}}$=0.702,\unboldmath ~20 lines)}
\nobreak
Fitting the continuum near 1750~\AA~was particularly difficult and
subjective, therefore the reality of the tabulated lines at 1765
and 1775~\AA~is suspect.  There are twelve \Lya~lines
identified in this spectrum.  We have identified the strong line at
1743.5~\AA~as \Lya~since there is no other plausible identification.
Note, however, that this line is well resolved at the 270 \kms
resolution of the FOS and is likely to be a blend.  No evidence for
any metal lines associated with this line was found in either the
complete or incomplete samples.  The line at $2000.8$~\AA~is quite
broad and located on the wing of a quasar emission line.  The
placement of the continuum fit might have artificially inflated the
significance of this line. There is no evidence of the expected
\hbox{Ly-$\beta$} line, but the formal uncertainty is high enough that
\zsearch~accepts the \Lya~identification.  The \Lyb~
line at $z=0.6727$ is considerably broader (FWHM$=3.0$~\AA) than its
matching \hbox{Ly-$\alpha$} line (FWHM$=1.95$~\AA). It is likely that the
\hbox{Ly-$\beta$} line is blended with a \hbox{Ly-$\alpha$} 
absorber at a redshift
of about 0.42.  The two Galactic Mg~II absorption lines bracket a
strong ``emission'' line that we are unable to identify (as either
real or an instrumental artifact, although we strongly suspect the
later).  This strong feature affects the measurement of the Mg~II
absorption and the inferred value for the equivalent with of the
Mg~II~\l2796.35 line (probably making it larger than its true value).
In addition to the Fe~II and Mg~II Galactic ISM lines identified in
the complete sample there is a line produced by Mg~I~\l2852.96 in the
incomplete sample with an equivalent width of $0.33\pm0.08$\AA~and
$SL$ of 4.1.  There is no evidence in the G160L spectrum for any
Lyman$-$limit or damped \hbox{Ly-$\alpha$} systems between 1200 and 2000~\AA.

\section{Results For Different Classes of Systems From the Census}
\label{sec-census}

The combined catalogue of absorption lines includes 3238 lines in the
complete sample of lines, found in the spectra of 78 quasars observed
with the higher resolution gratings. Eight of these 78 objects do not
have complete identifications for the observed lines (see
\S\ref{sec-identify}). We have also not yet constructed measured line
lists for the five BAL quasars presented in this paper and they do not
contribute to the total quoted above. Nine additional objects were
only observed with the G160L grating and do not contribute any lines,
other than some LLSs (which were not included in the total of 3238
lines), to the catalogue.  A rough indication of the redshift
distribution of the total path lengths of the survey for selected
types of absorption line systems is presented in Figure~3. The total
path lengths ($\Delta z$, ignoring for this paper variations in
equivalent width limit between the spectra) for Lyman$-$limit, damped
\Lya, \Lya, C~IV, and O~VI absorption systems are respectively 21, 49,
31, 44, and 17. Almost all of the survey observations (including the
objects observed only with the G160L grating, but excluding the five
BAL quasars presented in this paper), were included in constructing
the plots of the Lyman$-$limit and damped \Lya~system paths. For the
weaker \Lya, C~IV, and O~VI systems only the higher resolution data
(again, not including the five BAL quasars in this paper) were
included.  Mean or effective redshifts for Lyman$-$limit, damped \Lya,
\Lya, C~IV, and O~VI absorber surveys are respectively 0.76, 0.58, 0.72,
0.47, and 0.95. A more complete discussion of the effective path
length for \Lya~systems is included in Weymann~\etal~(1998).

While the notes on individual spectra include discussions of many
interesting individual absorption line systems, in the following
subsections we present brief descriptions of what the survey found
regarding selected classes of absorption systems.

\subsection{The \Lya, C~IV, and O~VI Absorption Systems}

The study of the evolution of \Lya~systems is a major focus of the Key
Project. Preliminary versions of our analysis have been presented
previously (\eg~CAT2; Jannuzi 1997, 1998). Our analysis based on the
entire combined catalogue is presented by Weymann \etal~(1998),
including discussion of the evolution of the number of \Lya~absorbers
with redshift and their distribution of rest equivalent widths.  The
combined catalogue includes 1,129 identified \Lya~lines.  A total of
1068 \Lya~lines was detected in the higher resolution spectra of the
objects whose line lists were completely identified (i.e., excluding
five of the six BAL quasars and the eight higher redshift quasars for
which the identification of the absorption lines are substantially
incomplete; see \S3~\& \S4).  In Figure~4 we show the observed
distribution in redshift of the catalogued \Lya~lines. 

Similarly, the evolution and nature of C~IV and O~VI systems are the
subject of separate papers and we will not discuss these systems in
detail in this paper. A total of 107~C~IV systems was identified in
the combined catalogue; 97 in the completely identified line
lists. Forty-one systems were identified that included absorption from
O~VI; 31 in the completely identified line lists. In Figure~4 we show
the observed distribution in redshift of the catalogued C~IV and O~VI
systems. The reliability of the C~IV and O~VI identifications is
discussed in \S3.4.

\subsection{The Incidence of Damped \Lya~Systems}
\label{sec-damped}

The \Lya~redshift path of the \HST~Quasar Absorption Line Survey is
comparable to previous surveys that investigated the incidence and
properties of damped \Lya~systems.  During the course of our survey a
redshift path of $\Delta z\approx49$ was surveyed with sufficient
sensitivity to find all damped \Lya~systems with
$N_{HI}>2\times10^{20}$ \hbox{cm$^{-2}$}.  The effective redshift of the
damped \Lya~survey was $<z>\approx0.58$.  Over the redshift path, one
damped \Lya~system was identified (in the spectrum of
PG~0935$+$416). This system has $z_{\rm{abs}}$=1.396 as determined
from a fit to the \Lya~line, independent of the algorithmic fit
discussed in the notes on individual spectra. The damped line and fit,
with $N(HI)=3.3\times10^{20}$ \hbox{cm$^{-2}$}, is shown in Figure~5. The
discovery of just one system at $z<1.65$ yields an observed number of
damped systems per unit redshift at $z=0.58$ of $(dN/dz)_{\rm
KP-damp}(z=0.58)=0.020$.  Poisson statistics apply to the
determination of the uncertainty in this measurement. Given our
observation of one system in a path length of 49, the minimum mean
number of damped \Lya~lines per unit redshift such that there is a
probability of 5\% that we see one line or more in our sample is 0.001
systems per unit redshift. The maximum mean number of damped
\Lya~lines per unit redshift such that there is a probability of 5\%
that we observe at most one line (i.e., 0 or 1 line) is 0.096 per unit
redshift . We adopt these values as our 95\% confidence boundaries on
the density per unit redshift of damped systems.

It is useful to compare our result with those from other studies on the
incidence of damped \Lya~at $z<1.65$.  Specifically, we can compare the
Key Project result to what is obtained from an interpolation of
results at lower and higher redshifts.  Rao \& Briggs~(1993) used 21
cm emission studies of gas in nearby galaxies to infer that, for
systems with $N(HI)>2\times10^{20}$ \hbox{cm$^{-2}$}, the local value of the
incidence of damped systems should be $(dN/dz)_{\rm
damp}(z=0)\approx0.015\pm0.004$. Moreover, the work of
Wolfe~\etal~(1995) suggests that the corresponding value at moderate
redshift is $(dN/dz)_{\rm damp}(z=2)\approx0.20\pm0.03$.  Based on a
linear interpolation of these results, the expected value for the Key
Project study is $(dN/dz)(z=0.58)\approx0.069$ marginally consistent
with our findings.

A number of previous direct studies of the incidence of damped \Lya~at
$z<1.65$ have been performed. The IUE-based survey of Lanzetta, Wolfe,
\& Turnshek~(1995) and the initial Mg~II-selected survey of Rao,
Turnshek, \& Briggs~(1995) reported $(dN/dz)_{\rm
IUE-damp}(z=0.8)\approx0.082$ and $(dN/dz)_{\rm
MgII-damp}(z=0.8)<0.12$, respectively.  However, more information
about the IUE-based survey and the initial Mg~II-selected survey is
now known in comparison to what was known at the time of their
publication.

In the case of the IUE-based survey, for which the quoted redshift
path was $\Delta z=49$ (virtually identical to the size of path length
available in our \HST~survey), five ``high-probability'' candidate
systems with \Lya~rest equivalent widths $>$ 10 \AA\ were assumed to
be damped, but it has been subsequently found that two of the
candidates (z$_{\rm abs}=0.519$ in Q1329+412 and $z_{\rm abs}=0.204$ in
Q2112$+$059) can be rejected on the basis of new FOS observations,
while one was ruled out at the time of publication (z$_{\rm abs}=0.399$ in
Q1318$+$290B).  Only one of the five high-probability candidate
systems is confirmed and the remaining high-probability system
($z_{\rm abs}=0.484$ in Q2223$-$052) must still be observed. The
confirmed high-probability candidate system is the same one found in
the Key Project ($z_{\rm abs}=1.369$ system in PG~0935$+$416).  At the
same time, a ``low-probability'' candidate system (with \Lya~rest
equivalent width $<$ 10 \AA\ as measured in an IUE spectrum) was found
to be damped ($z_{\rm abs}=1.014$ in Q0302$-$223), bringing the total
number of confirmed candidates in the IUE-based survey to two. While
this is a factor of two lower than was assumed at the time of the
Lanzetta \etal~(1995) publication, one high-probability candidate and
one low-probability candidate still need to be investigated to
formally finish the work on the IUE sample.  The success rate also
raises the concern that the redshift path over which damped \Lya~could
be found was somewhat over-estimated in the IUE survey.  Ignoring this
for now, based on the statistics of only two confirmations we find
$(dN/dz)_{\rm IUE-damp}(z=0.8)\approx0.041$, but note that this result
would increase if either of the two remaining candidates were
confirmed or if it was, in fact, determined that the redshift path was
over-estimated.  Using the $z=0$ result to interpolate to lower
redshift we find $(dN/dz)_{\rm IUE-damp}(z=0.58)\approx0.034$; a
result consistent with the findings of this paper.

Finally, the initial Mg~II-selected survey results of Rao
\etal~(1995) have now been considerably expanded by making
new FOS observations of Mg~II absorption-line systems (Rao \&
Turnshek~1998; Turnshek~1998).  Nine damped \Lya~lines in Mg~II
systems were uncovered and these new results indicate $(dN/dz)_{\rm
MgII-damp}(z=0.8)=0.10\pm0.04$.  Interpolating the Mg~II survey result
to $z=0.58$, again making use of the $z=0$ point, we find
$(dN/dz)_{\rm MgII-damp}(z=0.58)\approx0.077$, which is completely
consistent with the ``expected'' result obtained from interpolation
between the zero redshift and high redshift statistics. However, even
assuming that none of the proposed damped lines is really a complex
blend of weaker \Lya~lines, there is some possibility that the
$(dN/dz)_{\rm MgII-damp}$ statistic is over-estimated due to
gravitational lensing bias (Smette \etal~1997). But
Smette~\etal~(1997) point out that this same bias might affect the Key
Project and IUE survey statistics for damped \Lya, since both of those
surveys also utilized bright objects.  While the Mg~II-selected result
is nearly four times the Key Project result [$(dN/dz)_{\rm
damp}(z=0.58) \approx0.077$ versus $\approx0.020$], available
statistics for making a comparison are poor.  For example, if the
Mg~II-selected result is taken to be the true value, we estimate that
the probability of finding zero or one damped system in the Key
Project survey is $\approx$ 10\%, even though the most probable
observed number would then be four.  Clearly more work is required to
improve the accuracy of these statistics.

\subsection{Lyman$-$limit Systems in the Combined Catalogue}
\label{sec-limit}

In Table~6 we list the 16 candidate Lyman$-$limit systems observed in
the objects that were targets for the combined catalog.  Our analysis
of the evolution of LLSs has been previously presented in Paper~V.  Of
the 14 LLSs for which higher resolution spectroscopy is available to
search for associated metal lines (for all but the systems in 4C~19.34
and PKS~1055$+$20), all but the system in the spectrum of MARK~132
have identified metal lines associated with the LLS. In the cases of
the systems in the spectra of PG~0935$+$416, MARK~132, Q~1101$-$264,
S4~1435$+$63, and PG~1715$+$535 the line identifications are quite
incomplete and the association of additional metal lines with the LLS
(or in the case of MARK~132, any metal lines) can not be ruled
out. For eight of the 13 LLSs with detected metal lines the systems
are ``extensive'' (absorption systems with four or more observed metal
ions). The detection of strong metal lines, often part of an
extensive system including lines from both low and high ionization
states, is consistent with what was found in CAT1 and CAT2 regarding
the correspondence between LLSs and metal line systems and provides
additional circumstantial evidence that LLSs are likely to be
associated with galaxies (see CAT1 and CAT2 for further discussion).

\subsection{Extensive Metal Line Systems}

\stepcounter{footnote}

There is a total of 37$^{13}$ extensive metal-line systems (absorption
systems with four or more observed metal ions) listed in the tables of
identified lines (in CAT3, Table~3) in the combined catalogue of
absorption lines.  \footnotetext{Strictly, this number is a lower
limit on the total number of such systems in the combined catalogue,
as a small number of additional systems have candidate metal ion lines
that might be blended with lines whose primary identification is with
another system. Such lines are listed in the notes on individual
objects. If such lines were added to the count of metal ions
associated with each system, some would meet the requirement of four
detected metal ions and be called ``extensive''.  As we are only using
the definition to give a qualitative impression of the abundance of
metal line systems with numerous detected lines, we will restrict our
discussion to systems for which at least four metal ions are included
in the line identification tables.}  Twenty-seven of these systems
were added by the analysis presented in this paper.  These systems are
potentially valuable laboratories for understanding galaxies at
intermediate redshifts if they can be associated with a particular
galaxy (\eg~the case of the galaxy in the field of PKS~2145$+$06,
Bergeron \etal~1994).  Of the 37 extensive metal line systems, four
(six) are associated absorption systems, i.e., have velocities within
3,000 (5,000) \kms of the quasar emission lines. Two additional
systems in the spectrum of \pg2302~might be associated with the quasar
although they are more than 50,000~\kms from the redshift of the
quasar (Jannuzi \etal~1996). Excluding these two systems as well as
the known associated systems, we are left with 29 extensive metal line
systems that appear to be intervening systems.  For 25 of the
extensive metal line systems the portion of the quasar spectrum that
would contain any associated LLS was observed. Of these 25 systems, 21
(19) are more than 3,000 (5,000) \kms from the quasar, i.e. are
intervening extensive metal line systems. Of these 21 (19), eight have
an associated Lyman$-$limit system. None of the associated extensive
systems have an accompanying LLS.

\subsection{BAL Quasars}

Extremely broad high-ionization absorption complexes with velocity
extents of 2,000 to 25,000~\kms occur in approximately 10\% of
radio-quiet quasars (Weymann \etal~1991).  This absorption is almost
certainly produced by material ejected from the source producing the
observed emission.  Such objects are broad absorption line quasars
(BALs; see Weymann \etal~1991 and Turnshek~1995 for reviews). In this
third catalogue we have included the reduced spectra of five BAL
quasars observed during our survey: UM~425, PG~1254$+$047,
PG~1411$+$442, PG~1700$+$518, and PG~2112$+$059.  This brings the
total number of BAL quasars in the combined catalogue to six
(including PG~0043$+$039).  There is nothing we are aware of in the
selection of targets for observation in the survey that would have
biased us toward including or excluding any of these objects in our
sample, even though most were known to be BAL objects before we
observed them.

\section{Summary and Companion Papers}
\label{sec-summary}

Using the FOS of the \HST~we have produced a large 
and homogeneously constructed database of quasar absorption line
systems at low to moderate redshifts. The database is suitable for the
study of many problems, and we have undertaken some of these
examinations including studies of the evolution of \Lya~(CAT2;
Weymann \etal~1998) and C~IV absorbers (CAT1; Sargent \etal~1998), the
evolution of Lyman$-$limit systems (Paper~V), the evolution of damped
\Lya~systems (\S5 of this paper), the clumping of \Lya~absorbers
around extensive metal line systems (CAT2; Jannuzi 1998; Jannuzi
\etal~1998), and the Galactic halo (Savage \etal~1993; Savage
\etal~1998).  Some especially interesting systems include low
redshift \Lya~absorbers suitable for extensive follow-up observations
(\eg~in the spectra of TON~28 and PG~1216$+$069), possibly physically
associated pairs of extensive metal line absorption systems (\eg~in
the spectrum of PG~0117$+$213), and systems known to be associated
with galaxies (\eg~in the spectrum of 3C~232).  The spectra of five
broad absorption line (BAL) quasars (UM~425, PG~1254$+$047,
PG~1411$+$442, PG~1700$+$518, and PG~2112$+$059) can be found in this
third catalogue, bringing the total number of BAL quasars in the
combined catalogue to six (including PG~0043$+$039).

\acknowledgments
This work was supported by contracts NAG5-1618, NAG5-3259, and grant
\hbox{GO-2424.01} from the STScI, which is operated by the A.U.R.A.,
Inc., under NASA contract NAS5-26555.  We thank Digital Equipment
Corporation for providing the DEC4000 AXP Model 610 system used for
portions of this work.  D.~Saxe and R.~Deverill created significant
portions of the software used to measure the properties of the
absorption lines.  We thank M.~Burbidge and the FOS GTO team for
providing the community with an excellent window into the UV. We thank
the staff of STScI that facilitated the planning and scheduling of our
observations, particularly R. Lucas, D. Golombek, and those who 
assisted us in obtaining the necessary flat-field calibration data
when it was available, T. Keyes and R. Bohlin.  We acknowledge useful
discussions over the course of the Key Project survey with E. Jenkins,
A. Laor, A. Lowell, L. Lu, D. Maoz, S. Morris, L. Spitzer, S. Rao, M.
Strauss, D. Weinberg, and B. Yanny.  We thank J. Barnes and M. Best
for advice on the proper use of Latex and the AAS Latex macros.
We thank the anonymous referee for a very careful reading of the paper
and comments that improved  the presentation of our results.

\newpage
\centerline{FIGURE CAPTIONS}
\bigskip\medskip

\vbox{
\noindent
FIG. 1.--- The distribution in Galactic coordinates (Aitoff
projection) of the 92 quasars observed as part of the {\it HST} quasar
absorption line survey. The solid circles, open circles, and open
diamonds are the 89 objects observed as part of the Key Project and
Bahcall GTO observations that meet the original selection criteria for
Key Project targets. The open circles indicate the location of 15
objects whose spectra include coverage of 1150 to 1600 \AA~observed at
higher resolution (FOS G130H observations). The open diamonds are the
nine objects that were only observed at low resolution, using the G160L
grating.  The three triangles (two of which are blended together)
indicate the location of the three Bahcall GTO targets whose redshifts
are too low to have met our original selection criteria (MARK 205,
PG~1411$+$442, and PG~1415$+$451), but which are included in our final
catalogue. PG~1411$+$442, indicated with an open triangle, was
observed with the G130H grating while the other two objects in this
set were not.  See \S2.1 for further discussion of the sample of
observed objects and Table~1 for the list of the observed objects.
}
\bigskip

\vbox{
\noindent
FIG. 2.--- aa through cn -- Ultraviolet spectra of 67 quasars obtained
with the Faint Object Spectrograph of the Hubble Space Telescope.  The
panels contain the combined spectrum of these objects obtained with
the indicated gratings (G160L, G130H, G190H, and/or G270H).  The short
vertical bars indicate the positions of absorption lines in the
complete sample.  The dotted line is the ``continuum fit''; see
\S\ref{sec-select} and Paper~II.  The lower line in each spectrum
plot is the 1$\sigma$ uncertainty in the flux as a function of
wavelength.  For the higher resolution observations, the
4.5$\sigma_{\rm det}$ equivalent width limit (\AA) for unresolved
lines is also shown as a function of wavelength.  Flat-field residuals
that are strong enough to have a significance level greater than
$4.5\sigma$ are marked with the symbol FF.  To be included in the
complete sample, a feature's equivalent width must exceed the value of
the 4.5~$\sigma_{\rm det}$ curve at the relevant wavelength [see
Eq.~(\ref{eq:MinCompleteSample})].  The objects are presented in order
of increasing Equinox 1950 right ascension, the same order in which
the spectra are discussed in the notes on individual objects section
of the paper, \S\ref{sec-notes}.
}
\bigskip

\vbox{
\noindent
FIG. 3.--- a through d -- Shown in these four panels are the number of
lines-of-sight as a function of redshift that could contribute to the
total path length observed for the listed type of absorption system
(Lyman$-$limit, Damped \Lya, \Lya, and C~IV).  In constructing this
figure no effort was made to determine what corrections are necessary
for the varying level of sensitivity of each FOS spectrum and as a
consequence this figure should only be used as a rough guide of the
redshift distribution of the survey path length.  Lines-of-sight that
were on the short-wavelength side of \hbox{a $\tau > 1$} Lyman$-$limit
systems were not included in the histograms.
}
\bigskip

\vbox{
\noindent
FIG. 4.--- a through c -- The observed redshift distributions of
absorption from H~I (\Lya), C~IV, and O~VI are shown in these three
panels. For the plots of the distribution of C~IV and O~VI doublets,
each observed doublet is only counted once. Absorption features due to
gas in our Galaxy were excluded.
}
\bigskip

\vbox{
\noindent
FIG. 5.--- Fit to the damped \Lya~absorption line in the spectrum of
PG~0935$+$416. The derived neutral column density for this system is
$N(HI)=3.3\times10^{20}$ \hbox{cm$^{-2}$}. This system is discussed further
in \S\ref{sec-notes}~\&~\S\ref{sec-damped}.
}

\end{document}